\documentclass{aa}
\usepackage{graphicx}
\usepackage{amssymb}
\usepackage{amsmath}
\usepackage{psfig}
\usepackage{epsfig}
\usepackage{natbib}

\newcommand{\diff}{{\rm\,d}}
\newcommand{\gsim}{\raisebox{-3.8pt}{$\;\stackrel{\textstyle >}{\sim}\;$}}
\newcommand{\lsim}{\raisebox{-3.8pt}{$\;\stackrel{\textstyle <}{\sim}\;$}}
\def\msun     {$M_{\odot}$}

\def\lsun     {$L_{\odot}$}

\begin{document}
        \title{Chemical evolution of the intra-cluster medium}

        \author{A.\ Moretti\inst{1} \and L.\ Portinari\inst{2} 
        \and C.\ Chiosi\inst{1}}

        \offprints{A.\ Moretti} 

        \institute{Department of Astronomy, University of Padua, 
			Vicolo dell'Osservatorio 2, I-35122 Padova, Italy\\
                \email {moretti,chiosi@pd.astro.it}
                \and Theoretical Astrophysics Center, Juliane Maries Vej 30,
			DK-2100 Copenhagen \O, Denmark \\
                \email{lportina@tac.dk}}

        \date{Received: May 2002. Accepted: ?}

\authorrunning{Moretti, Portinari \& Chiosi }

\titlerunning{Chemical evolution of the ICM}

        \abstract{The high metallicity of the intra--cluster medium (ICM) is 
generally interpreted on the base of the galactic wind scenario for 
elliptical galaxies. In this framework, we develop a toy--model to follow 
the chemical evolution of the ICM, formulated in
analogy to chemical models for individual galaxies. 
The model computes the galaxy formation history (GFH)
of cluster galaxies, connecting the final luminosity function (LF)
to the corresponding metal enrichment history of the ICM.
The observed LF can be reproduced with a smooth, Madau--plot like GFH
peaking at {\mbox{$z \sim 1-2$}}, 
plus a ``burst'' of formation of dwarf galaxies at high redshift.\\
The model is used to test the response of the predicted 
metal content and abundance evolution of the ICM to varying input
galactic models. 
The chemical enrichment is 
computed from ``galactic yields'' based on models of elliptical 
galaxies with a variable initial mass function (IMF), favouring 
the formation of massive stars at high redshift and/or 
in more massive galaxies. For a given final galactic luminosity, these model 
ellipticals eject into the ICM a larger quantity of gas and of metals 
than do standard models based on the Salpeter IMF.\\
However, a scenario in which the IMF varies with redshift as a consequence
of the 
effect of the the cosmic background temperature on the Jeans mass scale,
appears to be 
too mild to account for the observed metal production in clusters. 
The high iron--mass--to--luminosity--ratio 
of the ICM
can be reproduced only by assuming a more dramatic variation of the typical 
stellar mass, in line with other recent findings. The mass
in the wind--ejected gas is predicted to exceed the mass in galaxies by
a factor of 1.5--2 and to constitute roughly half of the intra--cluster gas.
}

\maketitle


\section{Introduction}
\label{sect:introd}

The popular galactic wind (GW) scenario for elliptical galaxies, 
introduced by \citet{L74} to account for their photometric properties,
predicted as a side effect the pollution of the intra--cluster medium (ICM) 
with the chemical elements produced and expelled by individual 
galaxies \citep{LD75}. 
Metals in the hot ICM were in fact detected soon afterwards \citep{M76,S77}.

Iron is generally used as tracer of the 
overall metallicity, being the best measured element in the hot ICM.
Typical iron abundances in the ICM are around 0.2--0.3 solar 
\citep{R97, F98, Mus99, F2000, Fin01}.
The ICM seems to be also rich in $\alpha$-elements, for which ASCA
provided more firm estimates;
data 
by \citet{Mu96}
yield $\rm [{\alpha \over Fe}]\simeq 0.2 \pm 0.3$ or, in the case of oxygen,
$\rm [{O \over Fe}]\simeq 0.48^{+0.24}_{-0.55} $, which means, considering
the uncertainty,  a marginal over-abundance of oxygen relative to iron.
 \citet{Mu96} derive
$\rm [{Si \over Fe}]\simeq 0.37^{+0.17}_{-0.35}$, in agreement with
the analysis by \citet{F98}, who further suggest that $\rm [{Si \over Fe}]$ 
may increase with cluster richness/temperature.
\citet{IA97} and \citet{Ar97} argue that
considering the uncertainties and the revised value of the solar iron
abundance, abundance ratios are consistent with
solar. The first XMM and Chandra studies seem to indicate the same 
\citep{Buo2002}.

Recent studies have revealed a more complex distribution of metals in the ICM.
Gradients in the iron abundance have been detected with ASCA data 
\citep{F2000,Fin01, Whi2K},
and confirmed by BeppoSax data \citep{Ir2001,DeGra01};
in particular, sharp metallicity peaks in the central region seem to be
typical of cooling flows/cD clusters \citep{Fuk2K,Ir2001,DeGra01}.
[$\alpha$/Fe] ratios also seem to display a composite behaviour, 
increasing with cluster radius from
$\sim$ solar in the central regions to supersolar, typical type~II
supernova (SN), values in the outer regions \citep{F2000}. The existence
of gradients of metallicity and of abundance ratios is being
confirmed by the first XMM and Chandra results on a few clusters
\citep{Kaa2001, Tamu2001, Smi2002, John2002, Ett2002},
but no extensive cluster samples have yet been analyzed with these
satellites \citep{Buo2002}. It has been suggested recently that, 
in the presence of
abundance gradients, emissivity--weighted estimates of the average metallicity
might be higher than the true mass-weighted average, up to a factor of two
\citep{Pell01}.
We shall not discuss abundance gradients in this paper however, 
but only
the global average 
chemical
evolution of the ICM.

The source of such a large amount of metals 
in the ICM could be galactic,
as in the original prediction by Larson, or reside in Population~III 
pre-galactic objects \citep{WR78, Loe2001}. 
The distinct correlation between the iron mass in the ICM and the 
luminosity of elliptical and S0 galaxies,

\[ M_{ICM}^{Fe} \propto L_V^{E+S0}\]

\noindent
shown by \citet{A92}, seems to favour galaxies as the sites of 
production of the metals in the ICM.

Accepting that the metals in the ICM originated in the E and S0 galaxies of 
the cluster, two mechanisms exist to extract the newly
synthesized elements from the individual galaxies: the above mentioned 
GW and ram pressure stripping. Arguments have been given 
by \citet{R97} favouring the GW scenario, mostly based on the observation  
that the ``iron mass--to--luminosity ratio'' is roughly constant 
independently of cluster richness and temperature, while the ram pressure 
mechanism should be more efficient, extracting more metals for a given stellar 
content, in richer clusters.

Though the role of ram pressure stripping is still debated 
\citep{MB2000}, from here on we will limit to the GW scenario for the 
pollution of the ICM, bearing in mind that the addition of other mechanisms 
(metal production in pre-galactic objects and/or ram pressure stripping)  
would allow to inject even more metals into the ICM, further favouring the 
enrichment. In our models, GWs are powered by SN feed-back \citep{C2000},
as typically assumed in literature 
although AGNs have been suggested as an alternative source of energy input
\citep{R93,Rom2002}.

Over the past two decades, modelling the metallicity of the ICM has been the
subject of a great deal of studies that we summarize  briefly in the next
section. In this paper we present a model for the chemical evolution
of a cluster, in which galaxies eject part of their gas content by the GW 
mechanism, thus playing the role of stars in the classical models for the 
chemical evolution of the interstellar medium \citep{T80, M97}.

The plan of the paper is as follows.
Section~2 shortly reviews the problem of the gas and metal content of the ICM,
and previous studies in literature. In Section~3 we suggest that a
non-standard, variable initial mass function (IMF) for the stellar content
of a galaxy could improve upon our understanding of the problem, and discuss 
the corresponding galactic models.
Section~4 gives some simple estimates of the global properties of
clusters, as expected from galactic models with the variable IMF or with 
the standard Salpeter IMF; we introduce the concept of 
intra--cluster mass--to--light ratio (ICMLR) as a measure of the amount of
ICM gas.
In Section~5 we introduce our new model for the chemical evolution of a 
cluster as a whole, the underlying analogy between the inter-stellar and  
intra-cluster medium, the concepts of galactic formation rate
(GFR) and of galactic initial mass function (GIMF), and the model equations.
Section~6 contains the detailed discussion of a fiducial model,
calibrated on the observed ICMLR and the present-day luminosity function (LF) 
of galaxies as 
the key constraint for the GFR and GIMF; a comparison between results with
the Salpeter IMF and with the non--standard IMF is also made.
In Section~7, a set of models is presented with different galaxy formation 
histories. In Section~8 we present cluster models computed on the base of 
galactic 
models especially selected to reproduce the correct metal content of clusters.
In Section~9 we discuss the predicted [$\alpha$/Fe] abundance ratios 
in the ICM. Summary and conclusions are drawn in Section~10.

\section{Gas and metal in the ICM: previous studies}
\label{sect:gas_metal}

Various early studies investigated whether ``standard'' chemical models for 
galaxies can explain the amount of metals detected in the ICM 
\citep{V77,HB80,dey78}; by ``standard'' we mean a chemical model with the 
same physical ingredients (mainly, stellar IMF and yields) 
suited
to reproduce the Solar Neighborhood. Amidst these early studies, 
we mention in particular the one by \citet{MV88} as the first
attempt to link directly the metallicity of the ICM with the properties 
of the corresponding galaxy population. To this aim, the authors developed 
a modelling technique that has been widely adopted afterwards.
Basing on a grid of models of elliptical galaxies with GW, they assigned  
to any given galaxy of final stellar mass $M_*$, or equivalently of 
present--day luminosity $L_*$, the corresponding masses of gas and 
iron ejected in the GW, $M^{ej}_{gas}$ and $M^{ej}_{Fe}$.
Integrating these quantities over the observed LF,  
they calculated the total masses of gas and iron globally ejected by 
the galactic population in the cluster. Their main conclusions were:\\ 
(1) the iron content of the ICM can be reproduced with a standard 
Salpeter IMF in the individual galaxies;\\
(2) the global amount of gas ejected as GW is much smaller than the 
observed mass of the ICM, hence the ICM must be mostly primordial gas which
was never involved in galaxy formation.

Later on, this early success in reproducing $M_{Fe}^{ICM}$ turned out 
to be favoured by the low gravitational potential wells of model galaxies,
calculated only on the base of their luminous, baryonic component. Once the 
potential well of the dark matter halo is taken into account, the 
ejecta of SN~Ia hardly escape the galaxy and the metal pollution of the ICM 
by GWs is much reduced \citep{DFJ91a,MG95}. In this case, standard chemical 
models fail to reproduce the metal content of the ICM.
Some non--standard scenarios were thus invoked to solve the riddle, such as:

\begin{itemize}
\item[---]
a more top--heavy IMF than the Salpeter one, with logarithmic 
slope $x \sim 1.0$ rather than the standard value $x=1.35$ 
\citep{DFJ91b,MG95,GM97,GMl97,LM96};

\item[---]
a bimodal IMF with an early generation of massive stars heavily 
polluting the ICM, followed by a more normal star formation phase producing 
the stars we see today \citep{A92,EA95}.
\end{itemize}

These models, where SN~II from massive stars play the main role in the 
metal pollution of the ICM, were further supported by the enhanced 
abundances  of $\alpha$--elements with respect to iron detected with 
ASCA \citep{Mu96}.

Just as in the early work by \citet{MV88}, most authors
conclude that GWs cannot account for the huge amount of gas present
in the ICM, which is  2--5 times the mass in galaxies \citep{A92}.
The ICM must then consist, 50 to 90\%, of primordial gas.

\citet{T94}, on the base of the steep slope of the LF at the low
luminosity end observed in clusters, suggested that all the intra--cluster
gas could have originated in dwarf galaxies, since these are
numerous in clusters and they are expected to eject a large fraction of
their initial mass as GW, due to their shallow potential wells.
This suggestion
was discarded by \citet{NC95} and by \citet{GM97},
who calculated detailed models of dwarf galaxies and related GW ejection
to show that galaxies cannot be the only source for the whole intra--cluster
gas, even in the case of the steepest observed LF
(hence the largest contribution from dwarfs).

The models were further refined by \citet{MMC2K} who 
made use of one-zone and multi-zone GW models of elliptical galaxies 
and studied the dependence of the ICM abundances with  redshift. 
While the 
abundance ratios [O/Fe] are in both cases within the observational
uncertainties, the abundances [Fe/H] are 
very large and
require large dilutions by primordial gas. 

Very recently, \citet{Pip2002} showed that the Salpeter IMF might reproduce 
the observed metallicity of the ICM, provided 100\%  efficiency
of energy feed--back is adopted for SN~Ia. In this case, all the iron produced
at late times escapes into the ICM in a continuous wind/outflow.
The problem with this scenario, however, is that it inevitably predicts 
strongly sub--solar [$\alpha$/Fe] ratios in the ICM.

Finally, a recent attempt aimed at reproducing simultaneously the iron
abundance, the ratio [O/Fe], and the gas mass, was by \citet{C2000} who
made use of multi-zone models of elliptical galaxies and adopted a non-standard
IMF for their stellar content. More details on the models by \citet{C2000}
will be discussed in the next section as our present model stems 
from that work.

%
%

\section{A non--standard IMF}
\label{sect:non_stand_imf}
As reviewed in the previous section, a wealth of work in literature 
suggests that some 
non--standard scenario (or IMF) must be invoked to account for the 
metals in the ICM. 
We recall that a non--standard IMF has been suggested for elliptical galaxies 
also on the base of other, independent arguments:
\begin{itemize}
\item
a top--heavy IMF ($x \sim 1.0$) is better suited 
to reproduce the photometric properties of ellipticals \citep{AY87};
\item
systematic variations of the IMF in ellipticals of increasing mass 
might explain the increase of the mass-to-light (M/L) ratio with 
galactic luminosity, that is the so--called ``tilt of the Fundamental Plane'' 
\citep{L86,RC93,ZS96,C98}.
\end{itemize}
What physical reason may lead to a different IMF in different situations? 
From the theoretical point of view, a turn--over of the IMF at low masses 
is expected, related to the Jeans scale (thermal support)
and to the scale of magnetic support against gravitational collapse 
\citep{L98,PN2002}. The conditions of the ambient
gas may thus influence the cut--off of the IMF at low masses, and hence the
mass fraction of a stellar generation that is locked into ever-living, 
very low mass and low luminosity stars. A variation of this 
``locked-up fraction'' is crucial for the efficiency of the metal enrichment
produced by a stellar population \citep{T80}.

\citet{L98} suggested the following functional form of the IMF:
\[ \frac{dN}{d\log{M}} \propto M^{-x} exp \left( - \frac{M_s}{M}\right) \]
where $M_s$ is a typical mass scale related to the Jeans mass.
In brief, this IMF is a Salpeter power law down to a typical
peak mass
\[ M_p \sim \frac{M_s}{x} \]
below which there is an exponential cut-off. The peak mass
varies with the temperature and density 
of the parent gas as expected from Jeans' law:
\[ M_p \propto T^2 \, \rho^{-\frac{1}{2}}\]
In warmer and/or less dense gas, therefore,
the typical peak mass $M_p$ increases and the locked--up fraction is lower.
\citet{L98}, as well as \citet{C98,C2000}, pointed out that a minimum 
temperature for the star forming clouds is set by the cosmic microwave 
background, whose temperature increases with redshift, becoming higher 
than the typical temperature of present--day
molecular clouds at $z > 2$. At increasing redshift, the sole background then
produces an increasing Jeans mass and hence an IMF skewed toward massive 
stars. Other heating sources, like 
feed-back from massive stars or the UV background, and a reduced cooling rate
at low metallicities, may enhance the effect further.

From the observational point of view, the issue of the variation of the IMF
with ambient conditions is still open and very much debated.
In local studies, some authors underline that, within the uncertainties,
data are consistent with a constant IMF (Kroupa 2002), others find evidence
of variation, e.g.\ between cluster and field \citep{Mass98} in clouds of
different density \citep{Bri2002, Luh2003}, or between the disk and the 
halo of the Milky Way \citep{Cha2003}.

The IMF of the early star formation activity at high redshift is even 
harder to probe.
\citet{HF2001} find indirect evidence in the local halo for a typical stellar 
mass scale (Larson--like) increasing with redshift. A similar scenario has
been very recently advocated by \citet{Fin03} on the base of the different 
metal content in groups and clusters --- and the different typical redshift
of formation of their stellar content. Indications of a top--heavy
IMF at high redshift ($z>3-6$) has been also found for 
Lyman Break Galaxies \citep{Ferg02}.
All of this agrees with a number of recent theoretical studies suggesting 
that the first generations of stars were strongly skewed toward massive stars
\citep[][and references therein]{ABN2002, BCL2002, 
ChMonaco}.

In view of these results and theoretical arguments it is certainly 
a legitimate working hypothesis to consider an IMF with a physical 
dependence on the environment, through the typical Jeans mass.

\subsection{Galactic models with the PNJ IMF}
\label{sect:PNJmodels}
A behaviour resembling the one described above is predicted by the theoretical
IMF by \citet[][hereinafter PNJ]{PNJ}, which features a peak mass 
\[ M_p = 0.2 \, M_{\odot} \left( \frac{T}{10 {\rm K}} \right)^2  
\left( \frac{n}{1000 \, {\rm cm}^{-3}} \right)^{-\frac{1}{2}} 
\left( \frac{\sigma}{2.5 \,{\rm km\, s}^{-1}} \right)^{-1} \]
where $T$, $n$, and $\sigma$ are the gas temperature, number density, and 
velocity dispersion, respectively.
Although the physical derivation of the PNJ IMF has been sometimes questioned 
\citep{Sca98}, one can still adopt it 
in galactic models as a tentative recipe yielding the typical behaviour 
expected for the
Jeans mass; see \citet{C2000} for further discussion.

\citet{C98} developed chemo--thermodynamical models following
the thermodynamical evolution of the gas in an elliptical galaxy,
and the corresponding variations in the IMF according to the PNJ recipe.  
The characteristics and behaviour of these models as a function of galactic 
mass  and redshift of formation are discussed in full details in 
\citet{C98} and \citet{C2000}. Here, we briefly underline 
the qualitative trends with respect to (a) mass and (b) redshift of formation.
\begin{description}
\item[(a)]
At increasing galactic mass, the average density of the object $\rho$ 
decreases and the typical peak mass $M_p$ increases, 
yielding a lower locked-up fraction.
\item[(b)]
At increasing redshift of formation $z_{for}$, the temperature 
of the protogalactic gas increases, since it can never fall below 
the corresponding
temperature of the cosmic microwave background, $T \geq T_{CMB}(z_{for})$, 
which increases with redshift; hence, the peak mass $M_p$ is higher and the 
locked--up fraction is lower.
\end{description}
We remark here that the above mentioned trends are by no means drastic:
the peak mass $M_p$ hardly exceeds 1~$M_{\odot}$, and the 
``high $M_p$'' phase is limited to the early galactic ages; after the initial
stage, in fact, the system reaches a sort of thermodynamical balance, 
with the peak mass and the IMF settling on quite standard values.
The overall picture loosely resembles the bimodal behaviour suggested
by \citet{EA95}, with an early phase dominated by massive stars
followed by a more normal star formation phase producing the low--mass
stars we see today. However, in our models
the IMF naturally and smoothly varies in time following a physical 
prescription, rather than an imposed bimodal behaviour.

Though not long-lasting, the variations in the early phases suffice to
differentiate the resulting galactic models, making them
successful at reproducing many features of observed ellipticals, 
such as \citep{C98}: 
\begin{itemize}
\item
the tilt of the Fundamental Plane;
\item
the analogous of the ``G--dwarf'' problem, or the lack of a large population 
of low metallicity stars, detected
in the spectral energy distribution of ellipticals \citep{B94,Wor96};
\item
the high fraction of white dwarfs \citep{Bica96};
\item
both the colour--magnitude relation {\it and} the trends in 
$\alpha$--enhancement with mass {\it at the same time}, thereby
overcoming the well-known dichotomy between the ``classic'' and 
``inverse'' GW scenario \citep{M92, M94, M97}.
\end{itemize}
This last point is worth commenting further, as the modelling of the GW 
influences directly the predictions concerning the ICM. 
GW models of elliptical galaxies with a constant IMF (whether Salpeter
or more top--heavy) face the following puzzle.
The colour--magnitude relation suggests that GWs occur
{\it later} in more massive ellipticals than in smaller ones, 
so that star formation and chemical enrichment proceed longer 
and the stellar population reaches redder colours in more luminous objects. 
On the other hand, metallicity indices, if interpreted as abundance 
indicators, suggest that the [Mg/Fe] ratio increases with galactic mass;  
this requires GWs to occur {\it earlier} in more massive galaxies, where
SN~II should dominate the chemical enrichment to make the resulting
abundance ratios in stars $\alpha$--enhanced. 

This dichotomy between the so--called ``classic'' and ``inverse'' 
GW scenario, hampers predictions
of the metal pollution of the ICM, since two competing sets of GW models
are to be considered. It is therefore quite appealing, when we address
the chemical enrichment of the ICM, that the variable IMF scenario described
above can reproduce both observational constraints,
with a unique set of models.

\citet{C2000} first analyzed 
 the predicted
metal pollution of the ICM when galaxy models with the PNJ IMF are adopted.
To this purpose, he calculated multi--zone chemical models of ellipticals 
with the PNJ IMF.

The adoption of radial multi--zone models, rather than simple
one--zone models, has in fact important consequences on
the predicted enrichment of the ICM, as underlined first by
\citet{MMC2K}.
When the radial structure of an elliptical galaxy is considered,
with the corresponding gradients in density, colours etc., it turns out
that the GW does not develop instantly over the whole galaxy, but it tends
to set in earlier in the outskirts,
where the potential well is shallower, and later in the central parts.
This means that star formation and chemical enrichment proceed longer
in the centre than in the outer regions \citep{T98, MMC98}, 
and the GW ejected from different galactic regions is 
metal enriched to different degrees.

The models calculated by \citet{C2000} account for this effect
by dividing the galaxy into three zones: a central sphere where
star formation and metal production is most efficient and lasts longer;
an intermediate shell where the GW sets in earlier and the metal production
proceeds to a lesser extent; an outer corona where the gas is expelled
almost immediately, with virtually no star formation and chemical processing.
This behaviour is the combined result of the shallower potential well when
moving outward in the galaxy (as in standard models with a constant IMF)
and of the varying $M_p$ in the PNJ IMF when moving to outer, less dense
regions; see \citet{C2000} for a detailed discussion.

For the sake of comparison, analogous models with the Salpeter IMF
were also calculated, with mass range [0.18--120]~\msun, as suited 
to model elliptical galaxies \citep{T98}.

In all these models \citep{C98,C2000} the metal production
and recycling, and the resulting abundances of the GW ejecta, are followed
with the chemical evolution network developed by \citet{P98}.
Chemical yields of massive stars were derived from the stellar tracks
of the Padua group for the pre--supernova phases, and then linked to the
SN II models by \citet{WW95}, rescaled to the same core masses.
Yields for low and intermediate mass stars were taken from \citet{Mar96,Mar98},
however these are of minor importance in the present work: for the GW ejecta 
and the enrichment of the ICM, most important are iron and 
$\alpha$--elements, produced by supernov\ae. The chemical network also
includes type Ia SN, which are important iron contributors, adopting
the recipe by \citet{GR83} for the rate, and the ejecta from the W7 model by 
\citet{Thiel93}.
We refer the reader to the original papers for further details on the
chemical network. 

\begin{figure}
\begin{center}
\includegraphics[width=7.5cm]{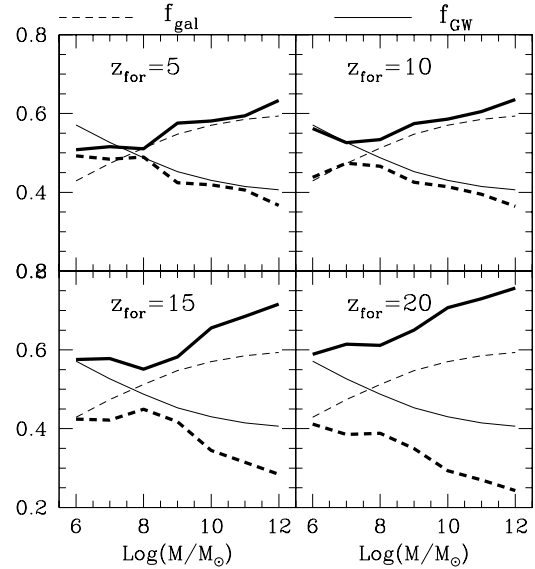}
\end{center}
\caption {Mass fractions in GW and in remaining galaxy as a function
of the initial (proto)galactic mass $M$, for four different redshifts 
of formation. {\it Thick lines}: galactic models with the PNJ IMF;
{\it thin lines}: galactic models with the Salpeter IMF.}
\label{fig:ejgas}
\end{figure}

\begin{figure}
\begin{center}
\includegraphics[width=7.5cm]{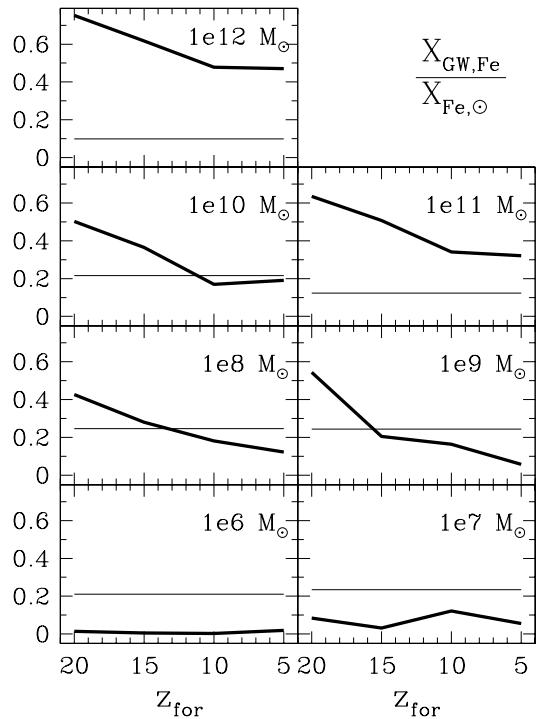}
\end{center}
\caption{Iron abundance (in solar units) of the GW, for galaxies
of initial (baryonic) mass $M$ as indicated in each panel, and as a function
of the redshift of formation. {\it Thick lines}: galactic models with 
the PNJ IMF; {\it thin lines}: models with the Salpeter IMF (redshift
independent).} 
\label{fig:ej_fe_cfr}
\end{figure}

\subsection{Galactic ejecta: PNJ vs.\ Salpeter IMF}
\label{sect:GWejecta}
For a better understanding of the results concerning the chemical evolution
of the ICM, let's first inspect the predicted GW ejecta of the galactic models
when the variable IMF or the Salpeter IMF are adopted in turn.

A (proto)galaxy of initial baryonic mass $M$ formed at redshift $z_{for}$ 
ejects a mass $E_{GW}(M,z_{for})$ of gas as GW,
while a mass
\[ R_{gal}(M,z_{for}) = M - E_{GW}(M,z_{for})\]
remains as the baryonic component of the galaxy we ``see today'', 
that is as the stars (and remnants) of the final 
galaxy.
For galactic models with the Salpeter IMF, the various quantities 
depend only on $M$ and not on  $z_{for}$,  as there are 
no temperature effects on the IMF in that case.
For the purpose
of the chemical evolution of the ICM, $R_{gal}$ can be viewed as a ``galactic 
remnant'', passively subtracting mass from further chemical/galactic
processing, analogous to stellar remnants in chemical models
of individual galaxies.
The 
gas shed by long--lived stars after the GW episode
is supposed to remain in the galaxy and not contribute further to the 
enrichment of the ICM.

In Fig.~\ref{fig:ejgas} we plot the mass fraction of gas ejected 
in the GW
\[ f_{GW} = \frac{E_{GW}}{M}\]
and the complementary mass fraction locked into the ``galactic remnant'', 
\[ f_{gal} = \frac{R_{gal}}{M} = 1 - f_{GW} \]
for galactic models with the variable PNJ IMF and for models with 
the Salpeter IMF ---
thick and thin lines, respectively. Mass fractions refer to
the total initial baryonic mass of the
(proto)galaxy. The amount
of ejected gas is larger in the case of the PNJ IMF, since less mass is locked
into low-mass stars thanks to the high $M_p$ in the early
galactic phases. In particular, with the PNJ IMF the mass ejected as GW 
is always larger than that locked up in the galaxy. The difference with 
the Salpeter case gets sharper
for larger (proto)galactic masses, and for higher redshifts of formation
(as already mentioned, models with the Salpeter IMF bear in fact no dependence
on the redshift of formation).
It is worth underlining here the following ``inverse'' behaviour of the
models with the PNJ IMF with respect to what is generally found
from models with a constant IMF. According to the general consensus,
larger galaxies store a larger fraction of their mass into stars and
eject a lower fraction of gas in the GW, while smaller galaxies,
due to their shallower potential wells, are more efficient in wind ejection.
This trend is indeed evident in the Salpeter galactic models of
Fig.~\ref{fig:ejgas}. The models with the PNJ IMF, on the other hand, show
quite the opposite behaviour: larger galaxies eject a larger fraction of their
initial mass in the wind and lock a lower fraction into stars, due to the 
higher peak mass that characterizes them in the early phases. This IMF effect
overwhelms their deeper potential wells, and the trend becomes stronger and
stronger with increasing redshift of formation.
There are in fact arguments 
advocating large baryon losses from galaxies
in general \citep{Silk02}.

The masses of iron and oxygen, respectively, ejected in the GW are
$E_{GW,Fe}$ and $E_{GW,O}$, so that the metal abundances of the GW
are given by:
\[ X_{GW,Fe} = \frac{E_{GW,Fe}}{E_{GW}} \qquad 
X_{GW,O} = \frac{E_{GW,O}}{E_{GW}}\]
The quantities $E_{GW}$, $E_{GW,Fe}$, $E_{GW,O}$ for a grid of values
of ($M$, $z_{for}$) are tabulated in Chiosi (2000, his Table~4,
detailing the contribution of the three different shells).

Fig.~\ref{fig:ej_fe_cfr} shows the iron abundance $X_{GW,Fe}$ 
in the gas ejected as GW,
again comparing the Salpeter IMF (thin lines) and the PNJ IMF case 
(thick lines).
In most cases, the galactic ejecta in the PNJ models are more metal--rich 
than in the Salpeter case, up to a factor of five or more in the case of the
most massive galaxies, and for high redshifts of formation. 
In the PNJ models, in fact, more galactic gas gets recycled
through massive stars, effective metal contributors, and less mass
gets locked into low--mass stars, before the GW occurs.

From the trends described above,  we expect that models of ellipticals
with the PNJ IMF predict a more efficient metal enrichment of the ICM,
and a higher fraction of its gas originating from GWs,
with respect to ``standard'' models. The first results in this respect
were discussed by \citet{C2000}.

\subsection{M/L ratio and IMLR of galaxies: PNJ vs.\ Salpeter}
\label{sect:GW_IMLR}
A popular way of measuring the efficiency of metal production 
of cluster galaxies is the so--called iron mass to light ratio (IMLR),
introduced by \citet{Ci91, R93}. This quantity is analogous to the
``global yield'' in chemical evolution models \citep{T80, Pag97},
defined as the ratio of the global amount of metals produced by 
a generation of stars to the mass 
that remains locked in remnants plus low--mass, ever living stars.
In the case of clusters, the mass locked in stars, i.e.\ in the stellar 
component of galaxies, is replaced by their luminosity (in the B--band), 
which is
directly measurable, while the estimate of the mass 
would be indirect
and would require some assumption about the M/L ratio of 
cluster galaxies. \citet{R97, R2000} estimates, for the IMLR of the ICM:
\[ \frac{M_{ICM}^{Fe}}{L_B} = (0.02 \pm 0.01) \, h_{50}^{-\frac{1}{2}} \]
where $M_{ICM}^{Fe}$ is the mass of iron in the ICM gas,
$L_B$ is the global luminosity of cluster galaxies, and $h_{50}$ is the Hubble
constant in units of {\mbox{50 km sec$^{-1}$ Mpc$^{-1}$}}. 
The IMLR appears to be quite constant among rich clusters \citep{R97}.
More recent estimates of the IMLR taking into account the existence 
of radial gradients of iron abundance in the ICM and considering gas masses, 
iron masses and luminosities within the same radius,
have slightly lowered the estimate \citep{F2000, Fin01}:
\begin{equation}
\label{eqn:IMLR}
 \frac{M_{ICM}^{Fe}}{L_B} = (0.01 - 0.015) \, h_{50}^{-\frac{1}{2}}
\end{equation}
These latter estimates have the advantage to be carried out consistently
over the same cluster volume. Metallicities are in fact 
typically measured at most within half of the virial radius 
\citep{F2000, Fin01, DeGra01,Ir2001}; measurements hardly reach 
$r_{500}$, which is typically 63\% of $r_{200}$, and in terms of cluster mass 
$M_{500} \sim 0.63 M_{200}$ as well \citep[see][]{Reip02}.\footnote{We 
indicate, as customary, with $r_{N}$ the radius
corresponding to an overdensity $N$ times larger than the average background 
density; $r_{200}$ is usually identified with the virial radius. $M_{N}$ 
indicates the mass enclosed within $r_{N}$.}
For the sake of the chemical enrichment what matters is the mass averaged 
metallicity,
obtained by convolving metallicity distributions with gas profiles
within the same radius, and the IMLR should also be evaluated from 
the gas, metals and galactic luminosity within the same radius 
\citep[as in][]{F2000,Fin01,Fin03}.
Since the observed metallicities probe only half of the
cluster mass, considering them representative of the whole
gas mass (derived from the gas profile extended out to the virial or Abell
radius) is a significant extrapolation, especially in view of recent
results on gradients: relaxed clusters with cooling flows 
systematically show iron abundance gradients (also besides the central peak)
and even the metallicity distribution in non--cooling flow clusters, 
though consistent with being uniform, is better fitted with
a negative gradient \citep{DeGra01}.

The IMLR could also be overestimated 
in the presence of a substantial intra--cluster 
stellar populations. Most studies limit this diffuse population to a 10--20\%
of the total stellar content, but values up to 40\%
have been recently suggested \citep{Arna02}.

Finally, we remind that red or NIR bands are a better probe of 
the actual star mass in galaxies,
for they are less sensitive to recent
sporadic star formation and more to the old underlying
population. Hence the IMLR would be a better probe of the real ``yield'' of
galaxies, if it were expressed in terms of the R to K band luminosity. This 
is becoming
possible nowadays, as cluster LFs in red or IR bands are presently becoming
available \citep{Mob03, Dri98, MT98, TM98, DPR99, AP2000, AC2002}; see in fact
\citet{Lin03}

While these developments are certainly interesting for the future,
here we will consider as our observational constraint the ``canonical'' 
value of the IMLR in the B band given by (\ref{eqn:IMLR}).

\begin{figure}
\begin{center}
\includegraphics[width=6.5cm,angle=-90]{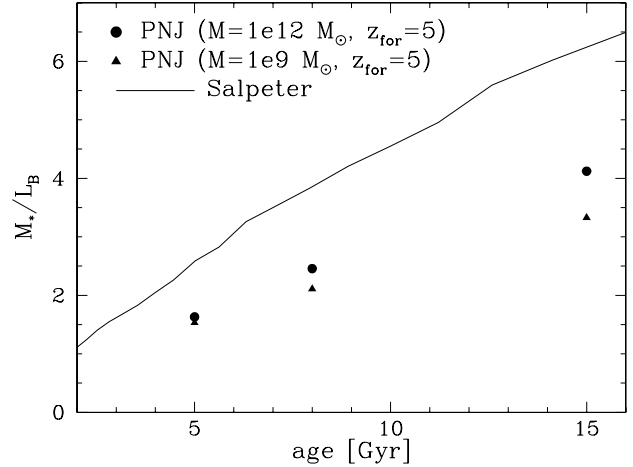}
\end{center}
\caption {The M/L ratio in the B band for living stars in the PNJ galaxies 
({\it dots and triangles}, two example models, see legend)
and in the Salpeter galaxies ({\it thin line}; mass limits for the Salpeter
IMF are [0.18--120]~\msun).}
\label{fig:MLB_stars}
\end{figure}

\begin{figure}
\begin{center}
\includegraphics[width=6.5cm,angle=-90]{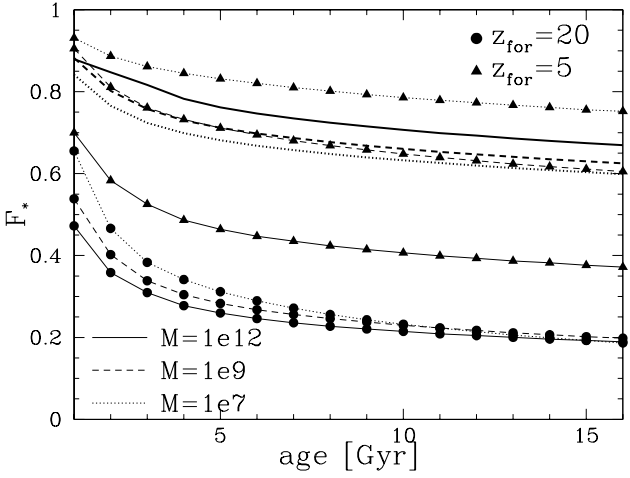}
\end{center}
\caption {Fraction of living stars within the ``galactic remnant'' $R_{gal}$,
decreasing with age in favour of remnants and ejected 
gas. PNJ models for three initial (proto)galactic masses and two redshifts of
formation shown for the sake of example (see legend). {\it Thick lines} with
no dots correspond to the Salpeter models.}
\label{fig:starfrac}
\end{figure}

\medskip
Before modelling the cluster and its IMLR as a whole, it is worth considering
what is the IMLR of the galactic wind of individual model galaxies:
\begin{equation}
\label{eq:IMLR_GW}
{\rm IMLR}_{GW} = \frac{E_{GW,Fe}}{L_B}
\end{equation}
comparing the PNJ IMF to the Salpeter IMF case. 
The GW ejecta $E_{GW,Fe}$ have been discussed above, now we discuss the 
B--band luminosities $L_B$ 
of our galaxies. A galaxy, or ``galactic remnant'' $R_{gal}$ left
over after the galactic wind, consists of three baryonic components: 
the living stars producing the galactic luminosity, the dark stellar 
remnants and the gas shed by stars after the GW. The first component 
progressively loses mass in favour of the other two; namely, within
$R_{gal}$ the fraction $F_*$ of living stars decreases with time
(Fig.~\ref{fig:starfrac}).

\begin{figure}
\begin{center}
\includegraphics[width=6.5cm,angle=-90]{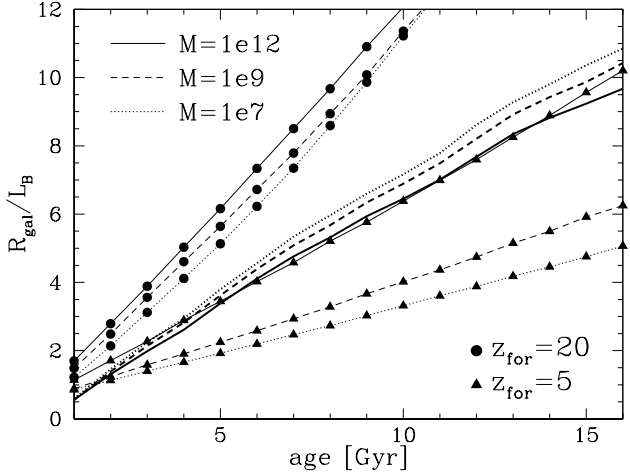}
\end{center}
\caption {M/L ratio of the global ``galactic remnant'' $R_{gal}$ 
(stars+remnants+gas ejected after the GW), as a function of age. Symbols as in 
Fig.~\protect{\ref{fig:starfrac}}.}
\label{fig:MLB_tot}
\end{figure}

Comparing the PNJ and the Salpeter models, two contrasting effects contribute
to determine the final luminosity of a galaxy. With the PNJ IMF, in the early 
galactic phases the typical stellar mass was skewed to higher 
values and less material was locked into low mass, very low luminosity stars.
As a consequence, the living stars are on average more massive 
and more luminous in the PNJ models than in the Salpeter models, so that 
their typical M/L ratio is lower (Fig.~\ref{fig:MLB_stars}).
On the other hand, with the PNJ IMF more mass went into remnants in the early,
top--heavy star formation phases, and the gas restitution fraction is higher 
(even after the GW) because of the lower number of ever--living low--mass 
stars; as a consequence, living stars often represent a lower mass fraction 
of the final galaxy in the PNJ models, especially for high redshifts of
formation and/or large masses (Fig.~\ref{fig:starfrac}).
Globally, the total M/L ratio of the galaxies, or ``galactic remnants''
made of stars+remnants+gas (ejected by stars after the GW), is shown in 
Fig.~\ref{fig:MLB_tot}. With respect to the Salpeter models, models with the
PNJ IMF show a composite behaviour: at low redshifts of formation 
($z_{for} \lsim 10$) they have lower M/L ratios, for the effect 
of having more luminous stars prevails; at high redshifts of formation,
they have higher M/L ratios, for the effect of the larger amount of remnants
prevails. 

\begin{figure}
\begin{center}
\includegraphics[width=6.5cm,angle=-90]{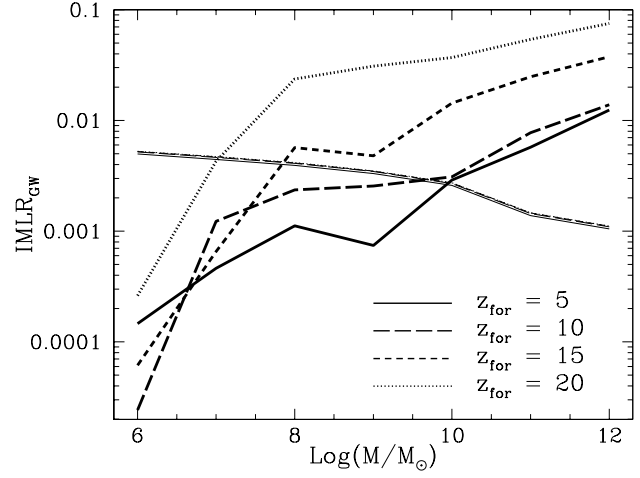}
\end{center}
\caption {The IMLR in the GW for model galaxies with the PNJ IMF ({\it thick
lines}) and the Salpeter IMF ({\it thin lines}).}
\label{fig:IMLRgal}
\end{figure}

Finally, Fig.~\ref{fig:IMLRgal} displays the IMLR$_{GW}$ 
of the galaxies (Eq.~\ref{eq:IMLR_GW})
as a function of the initial mass $M$ of the (proto)galaxy and of its redshift
of formation $z_{for}$; thick lines are for galaxies with the 
PNJ IMF, thin lines are for the Salpeter case. Notice that 
IMLR$_{GW}$ depends on $z_{for}$ also for the Salpeter case, for although
$E_{GW,Fe}$ is independent of redshift with the Salpeter IMF, $L_B$ 
decreases due to increasing galactic age at increasing $z_{for}$
(the effect is however negligible between $z_{for}=20$ and~5).

The trend of IMLR$_{GW}$ with mass $M$ is opposite in the two cases,
reflecting the effects already commented upon with Fig.~1 and~2.
At small masses, models with the Salpeter IMF eject more gas and metals
in the GW, retaining less mass in stars and having a lower final luminosity;
their IMLR$_{GW}$ is thus higher than that of the PNJ IMF models. At large
masses, say $M$\gsim a few $10^{10}$~\msun, the behaviour is reversed
and it is the galactic models with the PNJ IMF that eject a higher amount
of metals relative to the luminous mass retained in the ``galactic remnant''.
For $z_{for}<5$, GW ejecta are the same as in the $z_{for}=5$ case, 
as the background
temperature has by then dropped to low values and there is no further evolution
on its basic effects on the IMF. Therefore for $z_{for}<5$, 
for both IMFs the corresponding IMLR$_{GW}$ changes (decreases)
just because of the effects of increasing $L_B$ at younger ages.

Most important for the sake of the chemical enrichment of the ICM, notice that
the Salpeter models for all masses have an intrinsic IMLR$_{GW}<0.01$, i.e.\ 
always lower than what is observed in clusters (Eq.~\ref{eqn:IMLR}). 
We can thus anticipated that model galaxies with the Salpeter IMF are 
incapable of reproducing the high IMLR observed in clusters.
On the contrary, models with the PNJ IMF at high masses and/or for high
redshifts of formation do reach IMLR$_{GW} \geq 0.01$.

To predict the IMLR in the cluster as a whole, we need to convolve 
the respective IMLR$_{GW}$ with the Galactic Formation History of the cluster;
we will see (\S\ref{sect:ModelA}) that models with the PNJ IMF yield  a higher
global IMLR, in better agreement with observations. This is due to the fact
that small galaxies, although dominating the LF in number, actually represent
a minor contribution in mass, so that the main role in the ICM pollution
is played by galaxies with $M \gsim 10^{10}$~\msun\ \citep[see also][]{Tho99}. 
For the latter type, 
galactic models with the PNJ IMF have in fact an IMLR$_{GW}$ 
of the order of 0.01--0.02, 
comparable to the typical value $\sim$0.01 for clusters.

\section{Global cluster results: the monolithic approach}
\label{sect:monolithic}
As mentioned in \S\ref{sect:gas_metal}, the most popular way to calculate
the metal content of the ICM of a cluster on the base 
of its population of elliptical galaxies, is to develop a grid 
of galactic models 
and integrate their GW ejecta over the observed LF, a method first introduced
by \citet{MV88}. In this approach, all the ellipticals in the cluster are 
assumed to be coeval, hence we will call this the ``monolithic'' approach.

In this section we adopt the monolithic method, with the PNJ and the 
Salpeter galactic models in turn. 
Our reference LF is the observed B--band LF by \citet{Tr98},
since it is very deep (down to magnitude {\mbox{$M_B=-11$)}} and since the
observed IMLR is also referred to the B--band luminosity (see 
\S\ref{sect:GW_IMLR}). This LF, displayed in Fig.~\ref{fig:LFtren}, is
a weighted mean of the LFs of 9 clusters at low redshift ($z<0.2$).
Besides showing the standard exponential cut--off at bright magnitudes
($M_B<-20$), characteristic of a \citet{S76} function, this LF also steepens
at the faint end ($M_B>-14$), that is in the regime of dwarf galaxies. 

The luminosity evolution of the ``galactic remnants'' is known as a function 
of the initial (proto)galactic mass $M$ and of the redshift of formation 
$z_{for}$ (\S\ref{sect:GW_IMLR}). As to the relation between the redshift 
$z_{for}$ and the corresponding age of the galaxy, we adopt a flat 
$\Lambda$ Cold Dark Matter ($\Lambda$CDM) cosmology with $\Omega_M$=0.3, 
$\Omega_{\Lambda}$=0.7 and 
{\mbox{$H_0 = 65$~km~sec$^{-1}$~Mpc$^{-1}$}. Once a value for $z_{for}$ is 
fixed, each luminosity bin of the LF corresponds to some initial
galactic mass $M$, and we sum the contributions of the different bins
to estimate the global amounts of ejected gas and metals.

The GW ejecta of the PNJ models are quite 
sensitive to the exact epoch (redshift) of formation of the individual 
galaxies, as discussed in \S\ref{sect:GWejecta}, and we perform
the exercise for $z_{for}$= 5, 10, 13, 15, 20. (The case $z_{for}$=13 is 
obtained by interpolation in the grid of galactic models). For the Salpeter 
IMF, instead, computing the case $z_{for}$=5 is sufficient because the GW
ejecta are fixed, and the age and luminosity 
differences for $z_{for} >$5 are negligible (Fig.~\ref{fig:IMLRgal}).

\begin{figure}
\begin{center}
\includegraphics[width=6.5cm,angle=-90]{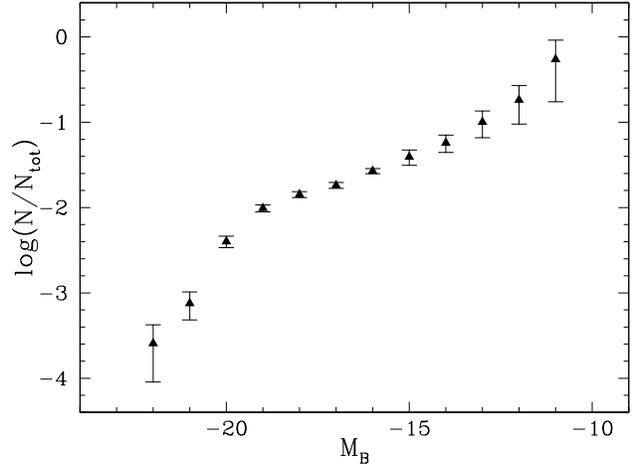}
\end{center}
\caption {The observed B--band luminosity function of cluster galaxies
by Trentham (1998), in relative frequency.}
\label{fig:LFtren}
\end{figure}

\begin{table*}
\begin{center}
\begin{tabular}{ c r|r r r r r |r r|r r r }
\hline

\medskip
IMF & \multicolumn{1}{c}{$z_{for}$} & \multicolumn{1}{c}{IMLR} & 
\multicolumn{1}{c}{$\frac{M_{gal}}{L_{tot}}$} & 
\multicolumn{1}{c}{$\frac{M_{GW}}{L_{tot}}$}
& \multicolumn{1}{c}{$\frac{M_{GW}}{M_{gal}}$} & 
\multicolumn{1}{c}{$\frac{X_{Fe}, GW}{X_{Fe, \odot}}$} & 
\multicolumn{1}{c}{$f_{SNIa}$} &
\multicolumn{1}{c}{[O/Fe]} & \multicolumn{1}{c}{$\frac{M_{ICM}}{M_{GW}}$} & 
\multicolumn{1}{c}{$\frac{M_{ICM}}{M_{gal}}$} & 
\multicolumn{1}{c}{$\frac{X_{Fe}, ICM}{X_{Fe, \odot}}$} \\

\hline
  PNJ &  5 & 0.007 &  7.2 & 11.1 & 1.54 & 0.358 & 0.53 & --0.08 & 3.3 & 5.1 & 0.11 \\ 
  PNJ & 10 & 0.009 &  8.2 & 13.0 & 1.59 & 0.376 & 0.54 & --0.07 & 2.8 & 4.5 & 0.13 \\ 
  PNJ & 13 & 0.018 & 10.6 & 21.1 & 1.99 & 0.477 & 0.47 &  +0.07 & 1.8 & 3.5 & 0.27 \\ 
  PNJ & 15 & 0.029 & 12.5 & 28.9 & 2.31 & 0.562 & 0.43 &  +0.16 & 1.3 & 2.9 & 0.44 \\ 
  PNJ & 20 & 0.065 & 17.4 & 50.6 & 2.91 & 0.714 & 0.36 &  +0.21 & --- & --- & ---  \\
\hline
 Salpeter &  5 & 0.002 &  8.6 &  6.2 & 0.72 & 0.143 & 0.65 & --0.09 & 6.0 & 4.3 & 0.02 \\ 
\hline
\end{tabular}
\end{center}
\caption{Results for integrated quantities for the cluster, in the monolithic 
approach. 
{\it 1$^{st}$ column:} IMF of model galaxies.
{\it 2$^{nd}$ column:} redshift of formation of galaxies.
{\it 3$^{rd}$ column:} global IMLR for the cluster.
{\it 4$^{th}$ column:} global M/L ratio for cluster galaxies (ratio between
		       the global mass in ``galactic remnants'' and the global
		       B--luminosity).
{\it 5$^{th}$ column:} ratio between the mass of gas ejected as GW and the 
		       global B--luminosity of cluster galaxies.
{\it 6$^{th}$ column:} ratio between the mass in gas ejected as GW and the 
		       mass in ``galactic remnants''.
{\it 7$^{th}$ column:} iron abundance with respect to solar in the gas ejected
		       as GW.
{\mbox{\it 8$^{th}$ column:}} fraction of the global iron production contributed
		              by SN~Ia.
{\mbox{\it 9$^{th}$ column:}} [O/Fe] ratio in the gas ejected as GW (and hence
		              in the ICM).
{\it 10$^{th}$ column:} ``dilution factor'' necessary to recover the observed
			ICMLR=$M_{ICM}/L_{tot}$=37~\msun/\lsun (see text).
{\it 11$^{th}$ column:} ratio between the total mass of ICM gas and the mass
		        in galaxies.
{\it 12$^{th}$ column:} average iron abundance in the ICM gas.
}
\label{tab:monolithic}
\end{table*}

Results for global integrated quantities with the PNJ and the Salpeter models
are listed in Table~\ref{tab:monolithic}. The 3$^{rd}$ column shows the 
global IMLR for the cluster. The Salpeter models fail to reproduce the 
observed IMLR by an order of magnitude or so. The PNJ 
models perform much better, with an iron production and hence an IMLR
increasing with redshift of formation, as expected from \S\ref{sect:GW_IMLR}.
For $z_{for}$=10--13 the IMLR falls in the
observed range (Eq.~\ref{eqn:IMLR}), while models with $z_{for} >$13 have 
too large an iron production with respect to observations.

\subsection{The contribution of Type Ia supernov\ae}
\label{sect:snIa}
At increasing redshift of formation, the PNJ IMF is more and more skewed toward
massive stars in the early phases, implying a lower and lower ratio of the number
of SN Ia vs.\ SN~II. Column 8 in Table~\ref{tab:monolithic} reports the fraction
of the global iron produced and ejected into the ICM due to SN~Ia; as expected,
it decreases with increasing redshift, from 55 to 35\%. 
The Salpeter IMF has a larger contribution (65\%) from SN~Ia than any case 
of the PNJ IMF. 

The same trends are reflected in the 
[O/Fe] ratio of the ejected gas (9$^{th}$ column in the Table),
increasing with redshift. Within the present uncertainties
about the [$\alpha$/Fe] ratio in the ICM (\S\ref{sect:introd}), all of the 
values for [O/Fe] in Table~\ref{sect:introd} are acceptable.
For the ``favoured'' cases (on the base of the IMLR) of 
$z_{for}$=10--13, the resulting [O/Fe] ratio is close to solar.

One concern when computing the iron production and ejection from galactic
models is the uncertainty in the rate of SN~Ia, due to the poorly known 
evolution of the progenitors.
As a consequence, one may wonder if the high iron
abundances in the ICM can be reproduced by adopting a higher rate
of SN~Ia rather than a top-heavy or non--standard IMF. However, the [O/Fe]
ratios we obtain indicate that the relative contribution of SN~Ia vs. SN~II
predicted for the ICM are grossly correct. 

With the Salpeter IMF and standard GWs models of elliptical galaxies,
the predicted contribution to the iron enrichment of the ICM
is short by 5--10 times with respect to the observed IMLR. 
If we were to compensate for this discrepancy by adjusting the SN~Ia 
rates to the required level --- or by assuming that later iron production
from SN~Ia escapes the galaxy after the main GW episode ---
the corresponding [O/Fe] ratios would decrease by 0.7--1 dex, by far
too low with respect to observations.

\subsection{The problem of dilution}
\label{sect:dilution}
Besides the amount of ejected metals as traced by the IMLR, an important
question concerns the global gas mass ejected from galaxies as GW: can this 
account
for the whole of the ICM, or conversely most of the ICM is primordial gas
that was never involved in galaxy formation? The observational constraint 
here is the global amount of ICM gas vs. the mass in galaxies; unfortunately,
this quantity is relatively poorly determined. In literatures, values can be
found in the range $M_{ICM}/M_{gal}$=2--20; a closer inspection reveals 
that this large range can be mostly imputed to different assumptions for
the $M/L$ ratios for galaxies. In fact, while the mass in the ICM can be 
estimated directly from
X-ray observations, the mass in galaxies is estimated from their global 
luminosity, usually in the B--band, and assuming a $M/L$ ratio. 
We remark here that for the chemical enrichment of the ICM,
what is meaningful is the baryonic mass in galaxies, not their
global mass or $M/L$ ratio inclusive of the dark halo component.
The real observable is $M_{ICM}/L_B$, and with respect to this quantity 
results in literature are much more homogeneous.

From the study of \citet{Dav90} of Hydra A
one finds $M_{ICM}/L_V = 33 \, h_{50}^{- \frac{1}{2}}$~\msun/\lsun;
the typical colours of ellipticals correspond to $L_V \sim 1.3 \, L_B$ and the
above entry translates into:
\[ \frac{M_{ICM}}{L_B} = 30 \, h^{-\frac{1}{2}} 
\, \frac{M_{\odot}}{L_{\odot}} \]
\citet{A92} assume $L_V \sim 1.2 \, L_B$ for the typical colours
of E and S0 galaxies, and consequently derive for their clusters 
$M_{ICM}/L_{V, (E+S0)} \sim (20-50)~h_{50}^{-\frac{1}{2}}$, with 
an average of 36~$h_{50}^{-\frac{1}{2}}$ \citep[see also][]{EA95}, 
corresponding again to:
\[ < \frac{M_{ICM}}{L_B} > \, \sim \, 30 \, h^{-\frac{1}{2}} \,
\frac{M_{\odot}}{L_{\odot}}~~~~~~{\rm with~a~range}~(17-42) \, 
h^{-\frac{1}{2}} \]
The analysis of the Coma cluster by \citet{W93} yields as well:
\[ \frac{M_{ICM}}{L_B} \sim 30 \, h^{-\frac{1}{2}} \, 
\frac{M_{\odot}}{L_{\odot}} \]
\citet{Cir97} estimate $M_{ICM}/M_{gal}$ for their sample under 
the assumption of a typical 
{\mbox{$M/L_V=8$~\msun/\lsun}} for elliptical galaxies; 
their tabulated values correspond to $M_{ICM}/L_V$ in the range 8.5--37.6 with
an average value of 21~\msun/\lsun. Taking again a typical 
$L_V \sim 1.3 \, L_B$ for ellipticals, and considering that Cirimele et~al.\ 
assumed $H_0=50$, their data imply:
\[ < \frac{M_{ICM}}{L_B} > \, \sim \, 19 \, h^{-\frac{1}{2}} \, 
\frac{M_{\odot}}{L_{\odot}} \]
In the sample by \citet{Rous2K}, the average over all the objects is:
\[ < \frac{M_{ICM}}{L_B} > 
\, \sim \, 31 \, h^{-\frac{1}{2}} \, \frac{M_{\odot}}{L_{\odot}} \]
while for the hot clusters only, a somewhat higher value is obtained:
\[ < \frac{M_{ICM}}{L_B} > \, \sim \, 44 \, h^{-\frac{1}{2}} \,
\frac{M_{\odot}}{L_{\odot}} \]
although it depends on the volume probed: $M_{ICM}/L_B$=27, 35 or 
{\mbox{44 $h^{-\frac{1}{2}}$~\msun/\lsun}}, for $r< r_{2000}, r_{500}$ or 
$r< r_{200}$
respectively, since the gas distribution is generally more extended than
the galaxy distribution.

\citet{Fin03} indicate a typical value of $M_{ICM}/L_B \sim$30 for hot 
clusters within 0.4~$r_{100}$, adopting $H_0$=70, or:
\[ \frac{M_{ICM}}{L_B} \, \sim \, 21 \, h^{-\frac{1}{2}} \]
In summary, data indicate a typical value of ``intra--cluster mass to light 
ratio''
\[ {\rm ICMLR} = \frac{M_{ICM}}{L_B} \, \sim \, 30 \, h^{- \frac{1}{2}} \,
\frac{M_{\odot}}{L_{\odot}} \]
with somewhat lower values in the samples by 
\citet[][]{Cir97, Fin03} (though still growing with radius in the latter)
and somewhat higher values in the sample by \citet{Rous2K}. 
Whether the fraction of hot vs.\ cold baryons, that is the efficiency of 
galaxy 
formation, increases noticeably with cluster temperature and richness
is on the other hand still a matter of debate 
\citep{Bryan2K, Bal01, Lin03}.

We favour $M_{ICM}/L_B$ as a constraint rather than 
$M_{ICM}/M_{gal}$, since the former is directly measured.
Just as the IMLR is supposedly a more straightforward estimate of the 
efficiency of metal enrichment than the ``classical'' yield $M_{Fe}/M_*$ (see 
\S\ref{sect:GW_IMLR}), the ICMLR
is a more direct constraint than the ``classical'' gas fraction $M_{ICM}/M_*$
because, like the IMLR, it is independent of {\it a priori} assumptions 
on the M/L ratio of galaxies. For our present favoured value of $h=0.65$, 
we regard ICMLR$\sim$37~\msun/\lsun\
as our constraint for the total amount of ICM gas in clusters. 

In the last three columns of Table~\ref{tab:monolithic} we give 
the dilution factors (i.e.\ the ratio between the total ICM gas mass 
and the mass provided by the GWs) necessary to recover the observed ICMLR, 
the corresponding ratio between ICM mass and mass in galaxies (the latter 
inclusive of stars, remnants and gas shed by stars after the GW) and the 
average metallicity in the global ICM. Typical dilution factors are in the 
range 2--3 for the PNJ 
models, and the ICM mass is typically 3--5 times the mass in galaxies.

In the case of $z_{for}=13$, that is the case with the best
results for the IMLR, roughly half of the ICM is expected to come from GWs,
its average metallicity is close to the observed characteristic value of
0.3 solar, and the ICM mass is 3.5 times the mass in galaxies.
Galactic models with a redshift of formation $z_{for}>15$ are ruled out since
they eject not only too much metals (as marked by the high IMLR) but also too
much gas in the GW to be compatible with the observed ICMLR. We notice also
that Salpeter models require a high degree of dilution (a factor of 6) and the
corresponding metallicity in the ICM gas is a factor of 10 too low, as expected
from the correspondingly low IMLR.

Our global M/L ratio for galaxies (including all the baryonic components,
living stars, remnants and gas shed by stars at late times) are in the range
7--12 for acceptable models with $z_{for} \leq 15$, and $\sim$8 for the 
Salpeter models. Correspondingly, the observed ICMLR=37~\msun/\lsun\ implies
that the mass in the ICM is 3--5 times the mass in galaxies. This is 
a factor of 2--3 lower than the widely quoted value of 10 derived by 
\citet{W93} for $h=0.65$. The difference just stems from their much lower
$M/L_B$=6.4~$h$, adopted on the basis of dynamical arguments; this is lower 
than what is expected for the typical stellar population in an old elliptical
galaxy (cf.\ the M/L ratio of Salpeter models). The difference 
is quite irrelevant for the problem discussed by \citet{W93}, that is the 
baryon fraction in clusters: the baryonic mass is dominated by the 
hot ICM gas mass anyways, and is not much affected by uncertainties 
in the mass in galaxies
 --- though the latter might not be as negligible, as sometimes assumed.
However, for the sake of the chemical enrichment of the ICM the effect is quite
crucial. Adopting the ICM--to--galaxy mass ratio by \citet{W93},
\citet{R97} estimates the ratio between the metals in the ICM and that locked 
in the stars of cluster galaxies to be $1.65 \, h^{-3/2}$, or $\sim 3$ for 
$h=0.65$.
If the M/L ratio in galaxies is higher, as required e.g.\ by stellar population
models for ellipticals, the ``metal balance'' is much less skewed toward the
ICM so that the amount of metals in the ICM becomes comparable to that
in galaxies.
This underlines the importance, for a consistent modelling of the chemical
evolution of the ICM, to adopt the observed ICMLR as a constraint rather than
some independently derived ICM--to--galaxy ratio relying on external 
assumptions on galactic M/L ratios. 

Finally, we remark that the actual ICM--to--galaxy mass ratio is important 
for the sake of explaining the ``entropy floor'' in low--T clusters, whether 
it requires strong supernova pre--heating or whether it is partly due to
the removal of low--entropy gas by galaxy formation 
\citep{Bryan2K,Bal01, Torna03}.
Interestingly, the recent cluster simulations by \citet{Val03} 
with star formation and self--consistent chemical evolution, suggest that 
the metallicities
of the ICM can be reproduced provided the IMF is top--heavy with respect to
Salpeter, and the corresponding mass in cold baryons (galaxies, or 
``star particles'' in the simulations) is large enough to favour the scenario 
of removal of low--entropy gas by \citet{Bryan2K}.

\begin{table*}
\begin{center}
\begin{Large}
\begin{tabular}{||c c c|c c c||}
\noalign{\vspace{5mm}}\hline\hline
\multicolumn{1}{||c}{}&\multicolumn{4}{c}{}&\multicolumn{1}{c||}{}\\
\multicolumn{1}{||c}{}&
\multicolumn{4}{c}{ ISM -- ICM ANALOGY  }&
\multicolumn{1}{c||}{}\\
\multicolumn{1}{||c}{}&\multicolumn{4}{c}{}&\multicolumn{1}{c||}{}\\
\hline
&&&&&\\
	    && ~~~~~\fbox{primordial gas}~~~~~	& 
~~~~~\fbox{primordial gas}~~~~~&& \\

	    && $\Downarrow$ 			& $\Downarrow$ && \\

& $\swarrow$ & {\normalsize SFR, IMF}		& {\normalsize GFR, GIMF} &
$\searrow$ & \\

	    && $\Downarrow$ 			& $\Downarrow$ && \\

ISM 	    && \fbox{stars}			& \fbox{galaxies} && ICM \\

	    && $\Downarrow$ 			& $\Downarrow$ && \\

& $\nwarrow$ & {\normalsize stellar yields} 	& {\normalsize GW yields} &
$\nearrow$ & \\

	    && $\Downarrow$ 			& $\Downarrow$ && \\

	    && \fbox{enriched gas}		& \fbox{enriched gas} && \\
&&&&&\\
\hline\hline
\noalign{\vspace{5mm}}
\end{tabular}
\end{Large}
\end{center}
\end{table*}

\section{A chemical model for the ICM  evolution}
\label{sect:toy-model}
The monolithic approach of \S\ref{sect:monolithic} assumes
all the ellipticals in the cluster to be coeval, with an age of 13--14~Gyrs. 
This is a somewhat na\"{\i}ve approach in the light of current understanding 
of structure formation in the Universe. In the hierarchical scenario the 
typical size of virialized objects increases at decreasing redshift. There 
are various reasons to argue that this scenario, developed for the 
gravitationally dominating, collisionless dark matter, might not hold for 
baryons, whose evolution at galactic scales is probably decoupled from that
of dark matter --- the angular momentum problem of disc galaxies in 
cosmological simulations, the observational evidence of early formation 
of elliptical galaxies, 
the colour--magnitude relation of both disc and elliptical galaxies 
suggesting the stellar populations in larger objects to be older at odds 
with expectations from the hierarchical bottom-up assembly, 
etc. Hence it is possible that
even in the hierarchical scenario, an old and roughly coeval stellar 
population of ellipticals can be accommodated, by suitable account of detailed 
baryon physics. Nevertheless, we explore now a picture in which the formation 
of cluster galaxies of different masses is redshift dependent; as the extreme 
alternative to the monolithic approach, we will assume that a galaxy forms 
at the epoch when the corresponding dark halo is predicted to virialize 
according to the Press-Schechter formalism.
We develop a chemical model following the epoch of formation 
of the individual galaxies in the cluster,
and evolve the overall system down to the present day, using 
the LF as a constraint {\it a posteriori}. 
This approach also allows, in principle,
a more self--consistent description of the chemical evolution of the cluster
and of its galactic population as a whole. 

An improved modelling of the evolution of the cluster, taking into 
account that its galaxies may form at different redshifts,
has been introduced by \citet{C2000}, who replaced the usual integration
over the LF with an integration over the \citet{PS74}
mass function at different redshifts. 
On the same line, we developed a global, self-consistent
chemical model for the cluster as a whole, 
following the
simultaneous evolution of all of its components: the galaxies, the primordial
gas, and the gas processed and re-ejected via GWs.
The approach is also somewhat
similar in spirit to the ``cosmic chemical evolution'' model by
\citet{Pei95},  aimed at calculating the global evolution of large volumes 
of the Universe, populated by a variety of star forming objects.

Our chemical model for clusters is developed in analogy with the usual
chemical models for galaxies. These latter
are schematically conceived as follows \citep{T80, Pag97}:
\begin{enumerate}
\item
the gas present in the system (usually starting from primordial composition) 
transforms into stars according to some prescribed Star
Formation Rate (SFR);
\item
stars are thus formed, distributed according to the adopted IMF;
\item
stars return part of their mass in the form of chemically enriched gas,
according to the so--called stellar yields (prescriptions derived from
stellar evolution and  nucleosynthesis);
\item
this chemically enriched gas mixes with the surrounding gas, causing
the chemical evolution of the overall interstellar medium (ISM).
\end{enumerate}
In a cluster, we are interested to model the chemical evolution of the
ICM, and the objects responsible for its enrichment are the galaxies,
via GWs.
In analogy with the above scheme, therefore, our chemical model 
for the cluster is conceived as follows:
\begin{enumerate}
\item
the primordial gas in the ICM gets consumed in time by galaxy formation 
according to some prescribed Galactic Formation Rate (GFR);
\item
at each time (redshift) galaxies form distributed
in mass according to a Galactic Initial Mass Function (GIMF), derived from the 
Press-Schechter mass function suited to that redshift;
\item
galaxies restitute a fraction of their initial mass in the form of
chemically enriched GWs, according to the adopted galactic models;
\item
this enriched gas mixes with and causes the chemical evolution of the 
overall ICM, which includes the amount of primordial gas not yet consumed 
by galaxy formation 
and the gas re-ejected by galaxies in the GWs up to the present age.
\end{enumerate}
The analogy between the two types of models is displayed in the scheme above.

Model equations parallel those of galactic chemical models, with the 
substitutions SFR $\rightarrow$ GFR, 
{IMF $\rightarrow$ Press-Schechter GIMF, 
stellar yields $\rightarrow$ GW yields. The main assumptions at the base 
of the model are as follows.
\begin{itemize}
\item
The model is one--zone, namely the cluster is treated as a single uniform
compound of gas and galaxies, where the metal abundance of the gas evolves
in time but is otherwise instantly mixed and homogeneous.\\
It is worth commenting here on how sensible one--zone
models are for structures, like clusters, that in current cosmological 
theories form out of hierarchical accretion of subunits.
We remind here that a one-zone chemical model contains 
no information about the spatial
distribution of its components. Therefore, at high redshifts our 
``model cluster'' can be considered simply as the sum of the subunits that
will later merge and form it, irrespectively of whether the cluster 
has in fact formed or not, as a single bound gravitational structure. 
The chemical model at high redshifts just describes the average properties
of the sum of the ``parent subunits'' of the cluster.

\item
The model is calculated assuming the Instantaneous Recycling Approximation
(IRA), that is assuming that galaxies eject the corresponding GWs instantly,
as soon as they are formed; this is a reasonable approximation as the
timescales for the onset of the GW are generally short, less than 1~Gyr.
The IRA could affect predictions at high redshifts, where
a time--span of a few $10^8$~yr corresponds to a sizeable gap in redshift; 
but up to redshift $z \sim 1$, where observational data on ICM abundances 
are available, the effect is minor.\\
Later on, the model might be improved in this respect, by taking into
account the actual delay between the formation of a galaxy and the time
when its GW is expelled; but as it is always better to start with 
the simplest possible assumptions, we adopt the IRA for the time being.\\
Notice however that the individual galactic models and their GW yields
are not calculated in the IRA, but with a detailed
chemical network taking into account the different, finite stellar lifetimes
at varying stellar mass 
(\S\ref{sect:PNJmodels} and references therein).

\item
\citet{Mor2001} investigated a set of pre-chosen, arbitrary GFR laws,
providing some indication as to what type of 
galaxy formation history is suited to reproduce the observed LF. 
An early burst of (dwarf) galaxy 
formation is needed to reproduce the steep faint end 
of the LF (see Fig.~\ref{fig:LFtren}); this burst should be followed 
by a more gentle GFR increasing 
in time (i.e.\ at decreasing redshift) up to a certain epoch, dropping after
that. Fig.~\ref{fig:closedmodel} shows such an example of suitable galaxy 
formation history (GFH) and corresponding LF. \\
In the present model, we try to include this kind of behaviour 
self--consistently, by means of a ``double infall'' scheme.

\begin{figure}
\includegraphics[width=8.5cm]{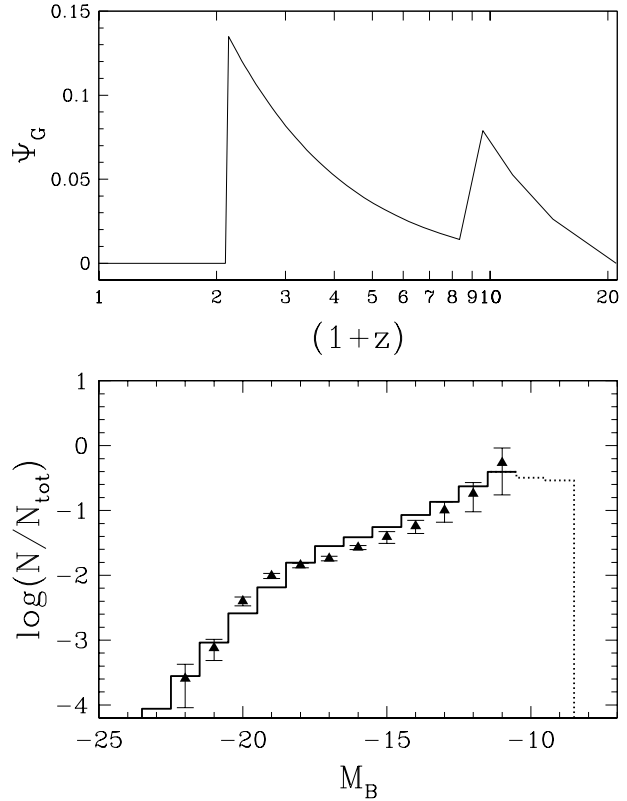}
\caption{An example of Galactic Formation History $\Psi_G$ [Gyr$^{-1}$]
as a function of redshift and corresponding Luminosity Function
(observational data by Trentham 1998). Model parameters are listed
in Table~\protect{\ref{tab:models}}.}
\label{fig:closedmodel}
\end{figure}

\item
Further simplifying hypotheses regard the history and the morphology 
of the galaxies contributing to the chemical evolution of the cluster: in
facts we have 
considered only elliptical non-interacting galaxies, since just early type  
galaxies 
seem to be responsible for the metal enrichment of the ICM 
\citep[see][]{A92}. Recent arguments about the baryon fraction 
in galaxies suggest however that early winds might be a general requirement
for all objects \citep{Silk02}, so in this sense the distinction between
elliptical or other galaxies might be of minor significance.
We do not explicitly take into account the possibility of 
mergers nor of episodic star formation after the GW.

\item
As to the cosmological parameters, which enter our model mostly through
the time--redshift relation, we adopt a flat $\Lambda$CDM cosmology with 
$\Omega_M$=0.3, $\Omega_{\Lambda}$=0.7 and 
{\mbox{$H_0 = 65$~km~sec$^{-1}$~Mpc$^{-1}$}}, corresponding to a present age 
of the Universe, or Hubble time, {\mbox{$t_H=14.5$~Gyr}}.\\
For the Press--Schechter Galactic IMF (see \S\ref{sect:GIMF}) we usually
assume $n=1.5$ as the index of the power spectrum.
\end{itemize}
%
\subsection{Basic equations}
At any time the cluster is assumed to consist of the following components:
\begin{itemize}
\item  
The gas with primordial chemical composition and total mass $M_{g,P}(t)$, 
out of which galaxies are  formed. 
\item 
The gas processed through galaxy formation and re--expelled by galaxies 
at the stage of galactic wind. Its total mass is $M_{g,W}(t)$. 
The wind--processed
gas contains several chemical species (carbon, oxygen, iron, etc.). We will 
explicitely consider here only oxygen and iron, with total masses denoted by 
$M_{g,O}$ and $M_{g,Fe}$, respectively.\\
We remark here that ``wind--processed gas'' in our models indicates
gas that has been processed through galaxies, not necessarily all of it
through the stars in the galaxies. In fact, the GW is made both
of metal enriched stellar ejecta and of pristine gas ``energized'' enough
by stellar feed--back to leave the galaxy as wind, without having ever
undergone actual star formation and stellar nucleosynthesis.
This pristine gas in the GWs should not be confused with what we label 
as ``primordial gas'' $M_{g,P}$ in
our cluster model, meant as gas that has never been involved in the
process of {\it galaxy} formation.
\item 
The galaxies, whose individual mass is the mass in stars and remnants left 
over after
the GW stage --- which occurs instantaneously as soon as the galaxy is formed,
in the IRA. The total mass in galaxies is denoted by $M_G(t)$.
\item
The total baryonic mass given by
\[ M_b(t) = M_{g,P}(t) + M_{g,W}(t) + M_G(t) \]
\end{itemize}
The Dark Matter component
does not enter model
equations as it does not intervene in the chemical evolution of the cluster.

As anticipated, for the primordial gas whence galaxies form we adopt a
(double) infall scheme; this scheme is meant to describe not quite 
the accretion of baryons onto the cluster region, but the accretion onto
the individual galaxies or, more precisely, the rate at which baryons become
available (through cooling etc.) for galaxy formation. As pointed out by
\citet{Pei95}, infall (or outflow) terms can be used even when dealing 
with large closed volumes, or with the whole Universe, as long as they 
represent gas exchange between the individual star forming objects
and the surrounding medium.
Owing to our double infall hypothesis, $M_b$ increases in time 
by accretion of primordial gas following the equation
\begin{equation} \label{eqn:mass_bar}
\frac{\diff M_b}{\diff t} \,=\, 
\left[ \frac{\diff M_{g,P}}{\diff t} \right]_{inf} \,=\, A \, t \, 
e^{- \frac{t}{\tau_1}} \,+\, B \, (t-t_0) \, e^{- \frac{t-t_0}{\tau_2} }
\end{equation} 
The first term represents a first, fast ($\sim 0.5$~Gyr) ``infall episode'' 
forming dwarf galaxies at high redshift; these objects are presumably
responsible for reionization and/or reheating of the intergalactic medium,
so that galaxy formation is temporarily halted.
Later on, gas cools down again
and galaxy formation progressively sets in again in a second, smoother
and prolongued ``infall phase'' represented by the second term (active
for $t > t_0$).
This double infall scheme was chosen so as to yield a shape of the
GFH suitable to reproduce the observed LF 
\citep[from a preliminary investigation by][see 
Fig.~\ref{fig:closedmodel}]{Mor2001}, 
as well as being reminiscent of current
evidence and theories about reionization at high redshift and of the observed
cosmic SFH of the Universe at lower redshift (the ``Madau--plot'').

The double infall scheme causes the baryonic mass of the cluster to increase
up to a final value $M_{b,T}$ at the present time $t_H$ (Hubble time,
corresponding to $z=0$). We adopt this final total baryonic mass $M_{b,T}$
as the normalization mass to which all values are scaled: 
\begin{equation} \label{eqn:mass_tot}
M_b(t_H) = M_{b,T} = 1
\end{equation} 
Hence, in Eq.~(\ref{eqn:mass_bar}) we can tune the timescales $\tau_1$ and
$\tau_2$ of the two infall episodes, as well as the amount of mass
involved in the second infall phase (set by the parameter $B$), while 
the parameter $A$ is fixed by the normalization~(\ref{eqn:mass_tot}).
\[ A \,=\, \frac{1 - B \tau_2 \left[ \tau_2 - (\tau_2 + t_H - t_0) \,
e^{-\frac{t_H-t_0}{\tau_2}} \right] }
{\tau_1 \left[ \tau_1 - (\tau_1+t_H) \, e^{-\frac{t_H}{\tau_1}} \right] } \]
With the adopted normalization~(\ref{eqn:mass_tot}), the masses
of the different cluster components ($M_{g,P}$, $M_{g,W}$ and $M_G$)
effectively become, as well as $M_b$,
mass fractions of the final baryonic mass $M_{b,T}$, and as such they are
dimensionless quantities. The equations governing their temporal evolution
are:
\begin{equation}\label{eqn:mass_gas_p}
    \frac{\diff M_{g,P}}{\diff t} = - \Psi_G(t) \,+\, 
	\left[ \frac{\diff M_{g,P}}{\diff t} \right]_{inf}
\end{equation}
\begin{equation}\label{eqn:mass_gas_n}
       \frac{\diff M_{g,W}}{\diff t} = \Psi_G(t) \, Y_G(t) 
\end{equation}
\begin{equation} \label{eqn:m_gal}
\frac{\diff M_{G}}{\diff t} \,=\, \Psi_G(t) \, [1 - Y_G(t)]
\end{equation}
The equations governing the masses of oxygen and iron 
$M_{g,O}$ and $M_{g,Fe}$ expelled into the ICM by GWs are:
\begin{equation}\label{eqn:m_gas_O}
       \frac{\diff M_{g,O}}{\diff t}= 
	\Psi_G(t) \, Y_{G,O}(t)
\end{equation}
\begin{equation}\label{eqn:m_gas_fe}
       \frac{\diff M_{g,Fe}}{\diff t} = 
	\Psi_G(t) \, Y_{G,Fe}(t)
\end{equation}
In the equations above,
$\Psi_G$ represents the Galaxy Formation Rate (GFR) --- the analogue of
the SFR in classical chemical models of the interstellar gas ---
while $Y_G$, $Y_{G,O}$ and $Y_{G,Fe}$ are the "yields" per galactic generation 
of total gas, oxygen, and iron respectively; all these quantities will be
specified in the following sections.

The total gas mass in the ICM is:
\[ M_g = M_{g,P} + M_{g,W} \]
and its oxygen and iron abundances are given by:
\[ X_O = \frac{M_{g,O}}{M_g} \qquad \qquad 
X_{Fe} = \frac{M_{g,Fe}}{M_g} \]
Finally, the initial conditions for our set of equations are
\begin{equation} \label{eqn:initial}
M_{g,P}(0)=0 \qquad M_{g,W}(0)=0 \qquad M_{G}(0)=0
\end{equation} 
together with
\begin{equation} 
M_{g,O}(0)=0  \qquad \qquad M_{g,Fe}(0)=0
\end{equation} 
The set of equations is integrated with a 4$^{th}$--order Runge--Kutta method,
dividing the integration time--span $[0-t_H]$ into 1000 timesteps.
The consistency of the solution is secured by the fact that at the final
epoch $t=t_H$ the sum of the different cluster components is:
\[ M_g(t_H) + M_G(t_H) = M_{b,T} = 1 \]
within a small error (of the order of $10^{-3}$).


\subsection{Galactic Initial Mass Function}
\label{sect:GIMF}

The GIMF adopted here stems from  the Press--Schechter (PS) mass 
function, originally devised to 
describe the collapse and formation of dark matter haloes from the primordial 
spectrum of perturbations.
Following e.g.\ \citet{LC94}, the mass function of collapsed objects
at redshift $z$ is:
\begin{align}\label{eqn:PS}
\frac{\diff f[z(t)]}{\diff \ln M}&=\left(\frac{2}{\pi}\right)^{\frac{1}{2}}
\left(\frac{n+3}{6}\right)\left(\frac{M}{M_*(z)}\right)^{\left(\frac{n+3}{6}
\right)}\nonumber\\
&\times\exp{\left[-\frac{1}{2}\left(\frac{M}{M_*(z)}\right)^{\left(
\frac{n+3}{3} \right)}\right]} 
\end{align}
where $n$ is the index of the cosmological power spectrum, $z$ is the redshift
corresponding to epoch $t$, and  $M_*$ is the cut--off mass, or characteristic
mass, at that redshift:
\begin{equation}
M_*(z) = M_{Nor} \times (1+z)^{-\frac{6}{n+3}}, \qquad M_{Nor} = M_*(z=0)
\end{equation}
Eq.~(\ref{eqn:PS}) yields the fraction of mass in dark matter haloes 
with mass $M$ at redshift $z(t)$, per logarithmic mass interval.
As discussed at the beginning of \S\ref{sect:toy-model},
for the purpose of the present study we assume that the same formula provides 
also the mass spectrum of the galaxies forming at redshift $z(t)$. 
For a fixed baryonic fraction in the Universe, typically $f_b=0.1$,
the ratio $M/M_*(z)$ in Eq.~(\ref{eqn:PS}) can be considered both 
as the ratio
between a dark halo mass and the corresponding ``characteristic mass'' 
at that redshift, and as the ratio between the corresponding baryonic 
masses. Namely, as the PS mass function is
expressed solely in term of a mass ratio, it can be used both as
a mass function of the dark haloes, and as a mass function of the 
corresponding baryonic components.
In the latter case, $M_*(z)$ becomes the baryonic mass of the (proto)galaxy
hosted by the ``characteristic halo'' at redshift $z$, or the typical 
mass-scale of (proto)galaxies at that redshift. In this context, $M_{Nor}$ 
is the mass of baryons contained in the typical dark halo mass collapsing 
at the present--day.

In our model the GIMF $\phi_G$ is defined as the number of galaxies 
per mass interval, given by:
\begin{equation}\label{phi_press}
\phi_G(M,t) = \frac{1}{M^2} \times \frac{df[z(t)]}{d\ln M}
\end{equation}
%
where the time dependence is due to $M_*(z(t))$ on the right--hand side.
 
While Eqs.~(\ref{eqn:PS}) and~(\ref{phi_press}) define the shape of the
GIMF, we need to define also the mass range $[M_i, M_u]$ within which galaxies 
can form at any redshift; in principle these mass limits can vary
with redshift. 

For the upper and lower mass limits we set, respectively: 
\[ M_u(z) =\gamma \, M_*(z)~~~~~~~~~~~~~~~~ M_u(z) =\frac{M_*(z)}{\gamma} \]
with $\gamma \simeq 100$; that is to say, at any given redshift galaxies
form with masses within two orders of magnitude the characteristic mass
at that redshift.
Fig. \ref{fig:xmlow} shows an example of the evolution of $M_*$, $M_i$ 
and $M_u$
as a function of redshift
(example with $M_{Nor}=3 \times 10^{13}~M_{\odot}$ and $\gamma=100$).

The resulting galactic mass function is finally normalized in mass, 
at any redshift, according to:
\[ \int_{M_i(t)}^{M_u(t)} M \, \phi_G(M,t) \, dM = 1 \]
so that $\phi_G(M,t)$ properly becomes the distribution function of the masses
of the galaxies formed at epoch $t$.

\begin{figure}
\includegraphics[width=6.5cm,angle=-90]{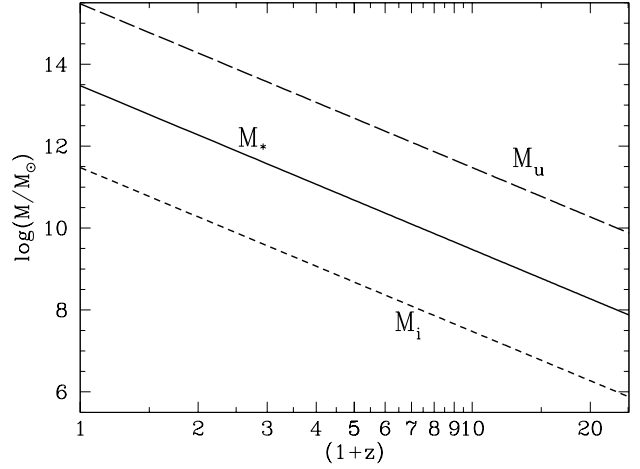}
\caption{Evolution of the characteristic mass $M_*$ (in this example,
with $M_{Nor}= 3 \times 10^{13} M_{\odot}$) and of the minimum and
maximum mass limits for the GIMF, $M_i=M_*/100$ and $M_u=100 \, M_*$.}
\label{fig:xmlow}
\end{figure}


\subsection{Global galactic yields}
\label{sect:Gyields}
Once the mass function of galaxies and its mass limits are specified
as a function of time (redshift), the global galactic yield of gas
and metals of each galactic generation can be determined 
out of integration of the GW yields of individual galaxies over 
the mass function.

Our galactic models, discussed in \S\ref{sect:GWejecta}, provide the mass 
of gas 
$E_{GW}(M,t)$, iron $E_{GW,Fe}(M,t)$ and oxygen $E_{GW,O}(M,t)$, ejected 
in the GW by a galaxy of given initial mass $M$
and formation epoch $t$.
The global quantities of gas,
iron and oxygen instantaneously recycled and re--ejected into the ICM 
by an entire galaxy generation formed at epoch $t$ can be determined as:
\begin{equation}\label{eqn:yields_gas}
 Y_G(t)=\int_{M_i(t)}^{M_u(t)} E_{GW}(M,t) \,\, \phi_G(M,t) \, dM
\end{equation}
\begin{equation}\label{eqn:yields_fe}
 Y_{G,Fe}(t)=\int_{M_i(t)}^{M_u(t)} E_{GW,Fe}(M,t) \,\, \phi_G(M,t) \, dM
\end{equation}
\begin{equation}\label{eqn:yields_o}
 Y_{G,O}(t)=\int_{M_i(t)}^{M_u(t)} E_{GW,O}(M,t) \,\, \phi_G(M,t) \, dM
\end{equation}
These are the quantities that enter 
Eqs.~(\ref{eqn:mass_gas_n}:\ref{eqn:m_gas_fe}) and which must be evaluated
at each timestep $t$ of the model, since all the quantities involved 
(GIMF, mass limits, GW ejecta) depend on the epoch $t$.
The mass that remains locked in the galaxies (``galactic remnants'',
see \S\ref{sect:GWejecta}) per galactic generation is:
\begin{equation}\label{eqn:yields_gal}
R_G(t)=\int_{M_i(t)}^{M_u(t)} R_{gal}(M,t) \, \phi_G(M,t) \, dM = 
1 - Y_G(t)
\end{equation}
entering Eq.~(\ref{eqn:m_gal}).
\subsection{Galactic Formation Rate}
\label{sect:GFR}

As to the Galactic Formation Rate, we adopt a simple \citet{S59} law
\begin{equation} \label{eqn:GFR}
\Psi_G = \nu \, M_{g,P}
\end{equation} 
where galaxies are assumed to form solely from the primordial gas component
with a rate that is directly proportional to the available ``fuel''.
With this law, the time evolution of the GFR is primarily set by the
``infall history'' of the primordial gas.

The Galaxy Formation History is supposed to start at epoch $z=20$
($\Psi_G(t)=0$ for $z > 20$), and it is halted 
when the characteristic mass forming at
that redshift, $M_*(z)$, becomes larger than that of the brightest
observed galaxy in the LF. In our simple ``hierarchical approach'', namely,
the formation of the brightest/largest galaxy corresponds to the
lowest typical redshift at which galaxy formation is active.

\section{A fiducial model}
\label{sect:ModelA}
The main observational constraint to calibrate the Galaxy Formation History
is the observed luminosity function. The LF by 
\citet[][Fig.~\ref{fig:LFtren}]{Tr98} 
shows the characteristic Schechter--like
exponential cut--off at bright magnitudes
($M_B<-20$), but it also steepens at the faint end ($M_B>-14$).
This large number of faint galaxies seems to require a prominent episode 
of formation of small galaxies at high redshift \citep[][and 
Fig.~\ref{fig:closedmodel}]{Mor2001}, a feature which inspired our present 
``double infall'' prescription.

The LF of cluster galaxies in our models is calculated by assigning to each
galaxy of ``remnant'' $R_{gal}(M,z_{for})$ the relevant 
M/L ratio for the age corresponding to its redshift of formation 
$z_{for}$ (see the discussion in \S\ref{sect:GW_IMLR}).

As to the normalization of the total number of galaxies in the cluster,
our models are calibrated also so that the final proportion of intra--cluster 
gas and galaxies respects the observational constraint of the 
ICMLR$\sim$37 (\S\ref{sect:dilution}), within a 20\%.

Here we present and discuss in detail our ``fiducial model'',
Model~A whose parameters and final results are listed in 
Table~\ref{tab:models}.
In this model we adopt as the characteristic mass for the GIMF 
(\S\ref{sect:GIMF})
\[ M_{Nor} = M_*(z=0) \,=\, 3 \times 10^{13} M_{\odot} \]
which is of the order of the characteristic mass of galaxy clusters nowadays
\citep{Gir98} once a factor {\mbox{$f_b=0.1$}} is applied to scale the
total mass to the mass of the sole baryonic component.

\begin{figure}
\includegraphics[width=0.35\textwidth,angle=-90]{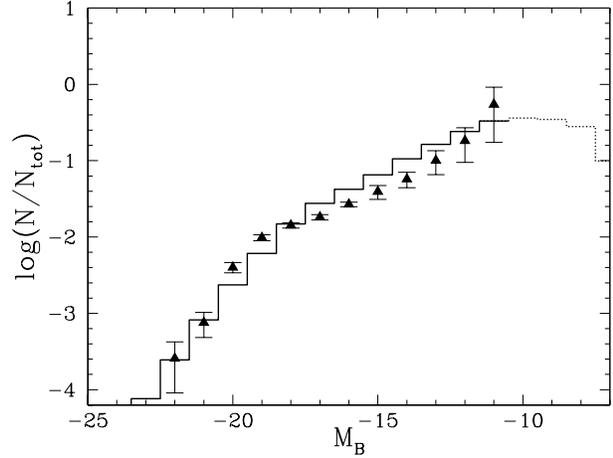}
\caption{Luminosity Function of cluster galaxies as predicted by our
``fiducial'' Model~A; triangles are the observational data by \citet{Tr98}.
Both the model LF and data are shown normalized to the same total (unit) 
number of objects in the observed range ($M_B$ between -22 and -11 mag).}
\label{fig:LFModA}
\end{figure}

\begin{figure}
\centerline{\includegraphics[width=0.35\textwidth]{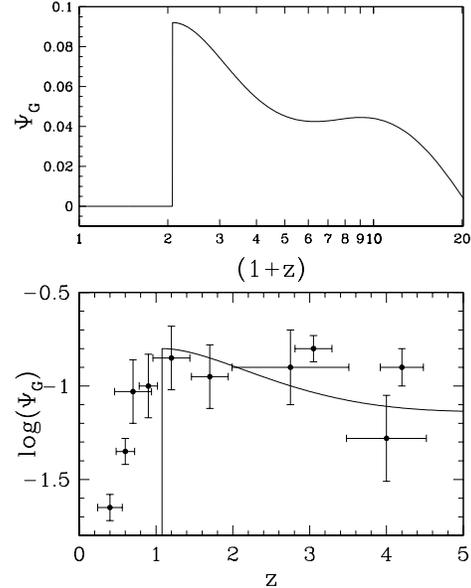}}
\caption{{\it Top panel}: GFH of our Model~A as a function of redshift;
$\Psi_G$ is expressed in mass fraction of the cluster per Gyr.
{\it Bottom panel}: comparison of the shape of our GFH with the ``Madau
plot'' \citep[data from][]{S99}; the theoretical curve $\Psi_G$
is arbitrarily normalized so that its maximum value reaches 
log($\Psi_G$)=--0.8, similar to the maximum in the data.}
\label{fig:madauModA}
\end{figure}

\begin{table*}
\begin{center}
\begin{tabular}{|c|c c|c c c c|c c c c c c|}
\hline
 & & & & & & & & & & & & \\
Model & IMF & $M_{Nor}$ & $\tau_1$ & $\tau_2$ & $B$ & $\nu$ & IMLR & 
$\frac{M_{gal}}{L_{tot}}$ & $\frac{M_{GW}}{M_{gal}}$ & 
$\frac{M_{ICM}}{L_{tot}}$ & $\frac{X_{Fe,ICM}}{X_{Fe,\odot}}$ & [O/Fe] \\
 & & & & & & & & & & & & \\
\hline
 & & & & & & & & & & & & \\
 A  & PNJ      & 3.e13  & 0.15 & 2.6 & 0.13 & 0.31 & 0.0058 & 5.6 & 1.56 & 
 32.8 & 0.096 & --0.061 \\
 AS & Salpeter & 3.e13  & 0.10 & 3.0 & 0.10 & 0.30 & 0.0013 & 7.1 & 0.72 &
 38.8 & 0.019 & --0.085 \\
 & & & & & & & & & & & & \\
 M1  & PNJ     & 5.e12  & 0.10 & 2.3 & 0.15 & 0.08 & 0.0042 & 4.4 & 1.53 & 
 29.9 & 0.076 & --0.064 \\
 A  & PNJ      & 3.e13  & 0.15 & 2.6 & 0.13 & 0.31 & 0.0058 & 5.6 & 1.56 & 
 32.8 & 0.096 & --0.061 \\
 M2  & PNJ     & 1.e14  & 0.10 & 0.9 & 0.80 & 0.25 & 0.0067 & 6.3 & 1.57 & 
 32.4 & 0.113 & --0.058 \\
 M3  & PNJ     & 3.e14  & 0.01 & 0.8 & 0.50 & 0.29 & 0.0074 & 7.0 & 1.58 & 
 35.7 & 0.114 & --0.037 \\
 & & & & & & & & & & & & \\
\hline
 & & & & & & & & & & & & \\
 AfY  & PNJ($z_{for}=15$) & 3.e13  & 0.10 & 3.4 & 0.09 & 1.50 & 0.0184 & 8.2 &
 2.26 & 40.6 & 0.247 & 0.163 \\
 M2fY & PNJ($z_{for}=15$) & 1.e14  & 0.10 & 1.6 & 0.35 & 1.25 & 0.0215 & 9.4 &
 2.29 & 39.2 & 0.299 & 0.164 \\
 & & & & & & & & & & & & \\
\hline
 & & & & & & & & & & & & \\
Fig.~\protect{\ref{fig:closedmodel}} & PNJ & 3.e13  &
\multicolumn{4}{|c|}{ input GFH as in Fig.~\protect{\ref{fig:closedmodel}} }
& 0.0061 & 5.8 & 1.56 & 30.7 & 0.108 & --0.060 \\
 & & & & & & & & & & & & \\
\hline
\end{tabular}
\end{center}
\caption{Parameters and results of the cluster models}
\label{tab:models}
\end{table*}

\begin{figure}
\includegraphics[width=0.35\textwidth,angle=-90]{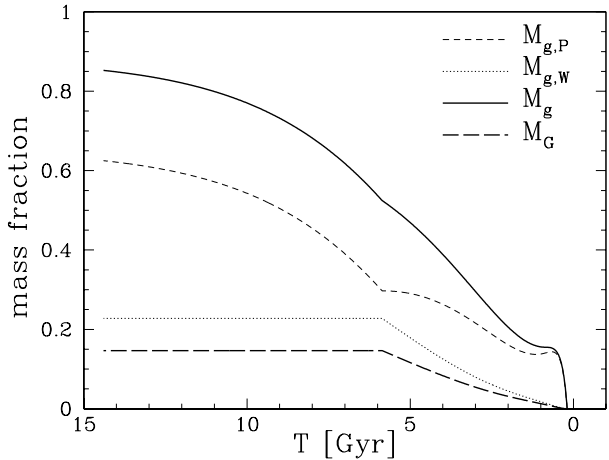}
\caption{Time evolution of the cluster components for Model~A}
\label{fig:compModA}
\end{figure}

\begin{figure}
\includegraphics[width=0.35\textwidth,angle=-90]{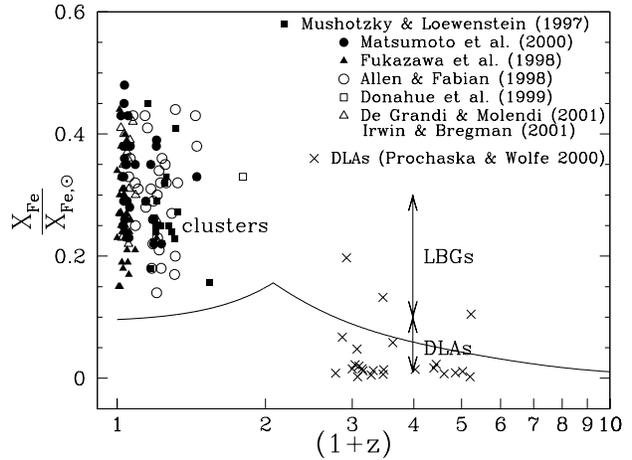}
\caption{Evolution of the iron abundance in the ICM. At high redshift
($z>1.5$) metallicities for Damped Lyman $\alpha$ systems and
Lyman Break Galaxies are shown for the sake of qualitative comparison
\citep[metallicity ranges for LBGs and DLAs at $z=3$ are from][]{Pet2K}.}
\label{fig:abundModA}
\end{figure}

With the adopted parameters the final ICMLR$\sim$33, the observed LF is 
reasonably well mimicked
(Fig.~\ref{fig:LFModA}) and the galactic GFH proceeds until $z \sim$1
(Fig.~\ref{fig:madauModA});
at lower redshifts, the typical galaxies forming would be more massive/bright
than observed in the LF. 
The steep, faint end of the LF is reproduced
by means of an early galaxy formation activity connected to the first 
infall episode (Fig.~\ref{fig:madauModA}, top panel); this early activity 
is short ($\tau_1 \sim 0.15$~Gyr)
and can be identified with the formation of the first objects responsible for
reionization. In fact, the second 
phase of galaxy formation 
(corresponding to the ``second infall'') slowly sets in starting from 
$t_0 = 0.8~$Gyr, in broad agreement with recent observations
suggesting that the reionization/reheating era took place at $z \gsim 6$
\citep{Dj2001, Beck2001}. Hence, we can loosely associate the 
trough in the galaxy formation activity at the transition between 
the two ``infall episodes'' (Fig.~\ref{fig:madauModA})
with the reheating of the intergalactic medium following the formation
of the first galactic objects; the subsequent slow cooling down of the
gas fuels the 
GFR at $z<6$.
The GFH eventually halts so abruptly due to the simplified prescriptions 
in the model (\S\ref{sect:GFR}); we remind also that only early--type
galaxies (E/S0) are considered in our cluster model, while in the Madau
plot for field galaxies, most of the SF at $z<1$ is due to spirals.

In Fig.~\ref{fig:LFModA}, the dotted part of the histogram represent 
the predicted extension of the LF to the range of galaxies fainter than
the observational limit. Although these small, faint galaxies dominate
in number, their contribution in terms of luminosity or stellar mass
is a negligible fraction of the total \citep[see also][]{Tho99}; 
in the Local Group as well, the numerous
dwarf galaxies play a negligible role in the global mass and luminosity
budget. In clusters, these faint objects might 
also have been disrupted and be nowadays dispersed as a diffuse intra--cluster
stellar component \citep[][and references therein]{Gregg98, Ferg98,
Arna02}.

\begin{figure*}
\begin{center}
\includegraphics[width=0.7\textwidth,angle=-90]{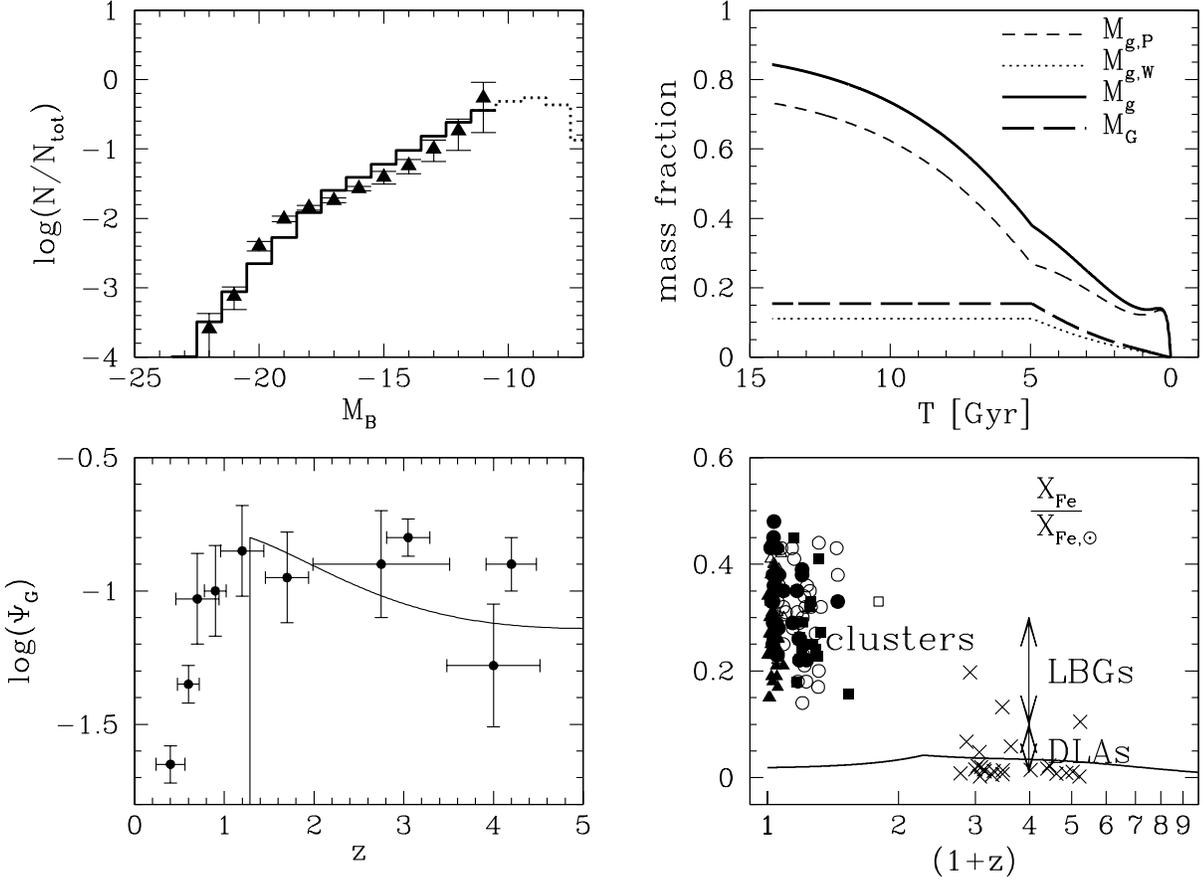}
\caption{Same as Figs.~\protect{\ref{fig:LFModA}--\ref{fig:abundModA}}, 
but for Model~AS based on the Salpeter IMF.}
\label{fig:ModAS}
\end{center}
\end{figure*}

Fig.~\ref{fig:compModA} shows the evolution of the mass fractions in the 
various cluster components: the mass in primordial gas, in wind--processed gas
and in galaxies. 
The ``double infall'' pattern can be recognized
in the evolution of the primordial gas component; within the first Gyr,
the early infall phase produces a minor peak --- first the gas mass 
increases due to infall, then it tends to decrease, being consumed by galaxy
formation; later on, again the increase due to infall competes with the 
consumption by galaxy formation, until the latter halts at
$z=1$.
At the end of the evolution, wind--processed gas is $\sim$1.5 times
the mass in galaxies, in agreement with qualitative expectations
from individual galaxies with the PNJ IMF (\S\ref{sect:GWejecta}).
The total mass in gas is about 6 times the mass in galaxies;
about 70\%  
of the ICM is primordial gas never involved in the galaxy formation process.

Fig.~\ref{fig:abundModA} shows the time evolution of metallicity 
(iron abundance) in the ICM; it peaks at $z \sim 1$, when the most
massive galaxies are formed and release their metal production,
then it decreases because primordial gas is still being
accreted onto the cluster, following our infall prescription, even after
galaxy formation and metal production are over.
The final  metallicity is low by a factor of 3 with respect to observations,
in agreement with the low overall IMLR ($\sim$0.006, Table~\ref{tab:models}).
This result is easily understood since most galaxies (in terms of mass 
involved)
form at redshifts $z \leq 5$, with a low corresponding
IMLR (see the discussion in \S\ref{sect:monolithic} and 
Table~\ref{tab:models}).

The predicted [O/Fe] ratio is roughly solar (Table~\ref{tab:models}).
If we were to compensate for the low iron production by increasing
the rate and iron contribution of SN~Ia by the required factor of 3,
the resulting [O/Fe] ratio would decrease by 0.5~dex, and become
too low with respect to observations, which indicate ratios between
solar and supersolar for the bulk of the ICM 
\citep{F2000}.

\subsection{Comparison to the Salpeter case}
For the sake of comparison, we calculate an analogous model
(Model~AS, with parameters listed in Table~\ref{tab:models}) 
adopting GW ejecta and ``galactic remnants'' as computed with the
Salpeter IMF. Fig.~\ref{fig:ModAS} is the analogue of Figs.~\ref{fig:LFModA} 
to~\ref{fig:abundModA} for Model~AS.
The LF is reproduced with a GFH qualitatively similar to that
of Model~A, but both the metal abundances in the ICM and the IMLR
(Table~\ref{tab:models}) are far from observational data.
Besides, in this case the mass of wind--ejected gas is comparable to,
or lower (70\%) than, the mass in galaxies and constitutes a minor fraction
(13\%) of the total final ICM.

Namely, with a Salpeter IMF galaxies produce by far too little metals per 
mass stored in galaxies. With the PNJ IMF things visibly improve
(Fig.~\ref{fig:abundModA}), still the expected metallicity is $\sim$3
times lower than observed.

\begin{figure*}
\begin{center}
\includegraphics[width=0.8\textwidth]{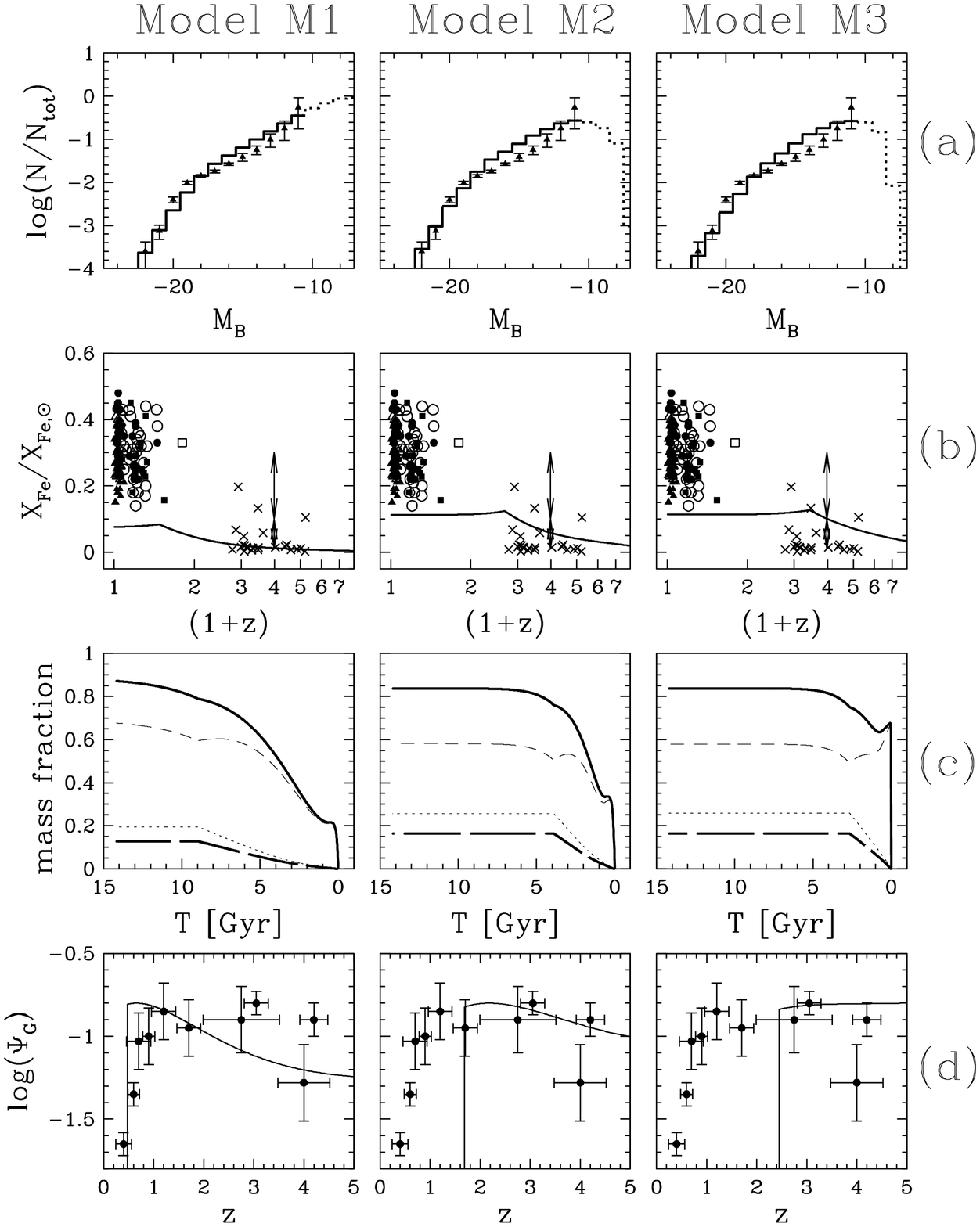}
\caption{Results for Models~M1, M2 and M3 which differ from our fiducial
Model~A in the characteristic mass $M_{Nor}$, increasing from M1 to M3
(Table~\protect{\ref{tab:models}}). (a) Luminosity Function;
(b) evolution of the iron abundance; (c) evolution
of the cluster components; (d) Galactic Formation History. 
Lines, symbols and data as in Figs.~\protect{\ref{fig:LFModA} 
to~\ref{fig:abundModA}}.}
\label{fig:Mnor}
\end{center}
\end{figure*}

\section{Varying the Galaxy Formation History}
\label{sect:varyingGFH}
In this Section we explore alternative models with ``slower'' or ``faster''
GFHs, by considering different characteristic masses
$M_{Nor}$ for the GIMF (\S\ref{sect:GIMF}). Qualitatively, increasing
$M_{Nor}$, 
the galaxies forming at a given redshift are intrinsically more massive
and luminous, so that to reproduce the observed LF the optimal GFH is ``anticipated'', or skewed to higher $z$.
Thus, $M_{Nor}$ is linked to the peak (and halt) of the corresponding GFH.
Similar effects could be obtained also by changing the spectral index $n$, 
for a shallower power--spectrum 
also induces the formation of intrinsically more massive galaxies 
at a given redshift, henceforth ``anticipating'' the required GFH. We explored
models with $n$ values different from the --1.5 adopted here, but we do not 
report them here for the sake of brevity,
as conclusions would be similar to those with varying $M_{Nor}$.


Models~M1, M2 and M3 are analogous to Model A, but for different
values of $M_{Nor}$; in each model, the ``secondary parameters'' governing the
GFH ($B$, $\tau_1$, $\tau_2$, $\nu$) are optimized so as to reproduce the 
observed LF and ICMLR.
Model~M1 has {\mbox{$M_{Nor}=5 \times 10^{12}$~\msun,}}
comparable to the characteristic mass at $z=0$ in the field, which in current
cosmologies is of the order of a few $10^{13}$~\msun\ (total mass,
dark matter included).
Models~M2 and M3  have a value of $M_{Nor}$ larger than in Model~A; 
their $M_{Nor}$ corresponds to the typical baryonic content of rich clusters,
with total mass of the order of $10^{15}$~\msun.

\begin{figure*}
\begin{center}
\includegraphics[width=0.8\textwidth]{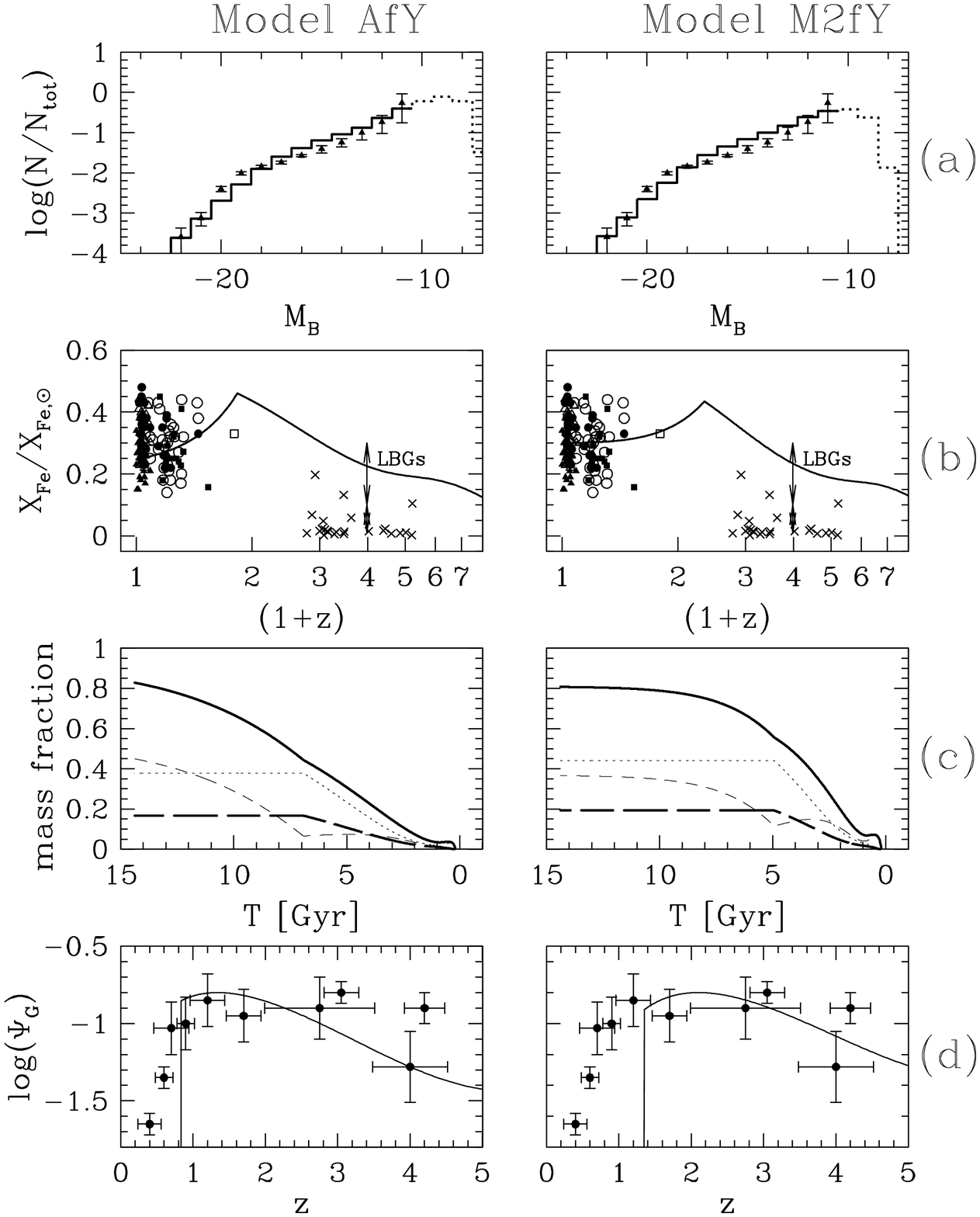}
\caption{Results for Models~AfYz15 and~M2fYz15, analogous to models A and M2
but with fixed galactic yields corresponding to the PNJ models with $z=15$.
Panels and symbols as in Fig.~\protect{\ref{fig:Mnor}}.}
\label{fig:fixYield}
\end{center}
\end{figure*}

As anticipated, at increasing $M_{Nor}$ (from M1 to M3), the GFH is skewed
to higher $z$ (Fig.~\ref{fig:Mnor}, d--panels); also 
the bulk of the metal production is correspondingly anticipated.
The main consequence is that
the global IMLR increases with $M_{Nor}$ (Table~\ref{tab:models}), 
since a given $L_B$ bin in the LF now
corresponds to older, dimmer and more massive galaxies, which eject both more
gas and more metals into the ICM, and have a higher characteristic IMLR$_{GW}$
(\S\ref{sect:GWejecta} and~\ref{sect:GW_IMLR}). 
Nevertheless, even in model M3 corresponding to the largest $M_{Nor}$,
the predicted IMLR is still low by a factor of 2 with respect to observations.
This is due to the fact that in all models most galaxies 
in terms of mass involved form anyways
at $z \leq$5, with a low characteristic IMLR, as was the case for 
model A. Correspondingly, the predicted metallicities are too low.

Hence, considering different GFHs --- in particular, ``anticipated'' GFHs
with respect to model~A, as in models M2 and M3 --- improves the situation
with metal production, but only slightly. The conclusion in the previous
section still holds, that if an IMF varying with redshift performs much better
than the Salpeter models, the metal production provided by the PNJ models
is still low.

\section{Choosing the IMF ``ad hoc''}
\label{sect:fixYield}
In the previous Sections we showed that a varying IMF like the PNJ IMF, 
skewed toward more massive stars (or more precisely, having a lower 
locked--up fraction) at higher redshifts, yields much better results 
for the predicted ICM metallicity, than the standard Salpeter IMF. 
Nevertheless, a still larger metal production seems to be required, 
by a factor of 2--3, than obtained with our PNJ models. In these models,
aside from the variation with galaxy mass due to density effects, the
peak mass of the IMF mainly varies with redshift because of the
CMB temperature (\S\ref{sect:PNJmodels}); apparently, a larger variation
is required to reach the required metal production.
Interestingly, this is in line with other recent results: both \citet{HF2001},
from the number of metal--poor stars in the halo of the Milky Way, and
\citet{Fin03}, from arguments related to the different yield in groups 
and clusters, suggest a typical peak--mass of the IMF increasing with redshift
faster than what expected from the pure Jeans effect of the CMB temperature.

We 
consider in this section some cluster models assuming 
an IMF that is more efficient in terms of gas and metal ejection in the
galactic wind. We achieve this by ``fixing'' the galaxy models 
to be the PNJ models with $z_{for}$=15: the GFH in the cluster still 
extends in time down to $z \sim$1 (or lower), but now we assume that galaxies
at any redshifts form with the properties (galactic yields and remnants)
of the PNJ models with $z_{for}$=15. Namely, the 
integrals~\ref{eqn:yields_gas} through~\ref{eqn:yields_gal} in 
\S\ref{sect:Gyields} are now computed always from the models with 
$z_{for}$=15, rather than with a $z_{for}$ running with the actual epoch $z$ 
of cluster evolution.
This is a simple artifact to mantain an IMF ``more efficient'' than the 
CMB--regulated PNJ IMF. Luminosities are of course
computed for the actual age $t(z)$ of the galaxies. 
The choice of the models with $z_{for}$=15 is {\it ad hoc} to obtain 
the correct final IMLR and metallicity (see \S\ref{sect:monolithic} and
Fig.~\ref{fig:fixYield}). The IMLR resulting from the ``hierarchical'' GFH
is somewhat lower than the monolithic case with the same $z_{for}=15$ because,
with galaxy formation extending down to redshift 1--2, galaxies are on average
younger and more luminous than in the monolithic case, lowering the typical
IMLR$_{GW}$ for a given mass of ejected iron.

In Fig.~\ref{fig:fixYield} we show the cluster models AfY and M2fY,
analogous to models A and M2 (i.e.\ with the same $M_{Nor}=3 \times 10^{13}$ 
and $10^{14}$, respectively) but calculated with ``fixed galactic yields'' 
as described above; parameters and results are listed in 
Table~\ref{tab:models}. In both cases the final metallicity and IMLR are
in good agreement with observations, the mass in ejected wind is $\sim$2.3 
times the mass in galaxies and constitutes roughly half of the whole ICM gas.
Notice also that, for these models reproducing the correct final ICM 
metallicity, the predicted metallicity at redshift 3--5 falls in the range
of Lyman Break Galaxies, which are in fact considered to trace the high redshift
counterpart of present--day massive galaxies in high density regions of the 
Universe, such as clusters.

In model~AfY, the GFH peaks at $z \sim$1 (similarly to the Madau plot), 
and correspondingly a noticeable peak in metallicity is predicted at the same 
redshift, for no further enrichment takes place for the primordial gas 
infalling onto the cluster later on. This fast metallicity evolution is not 
really observed --- 
although very few data points are available for redshifts $z>0.6$, where most 
of the evolution takes place.

A milder behaviour, somewhat more compatible with data, is found for
model~M2fY, having a characteristic $z=0$ mass $M_{Nor}$ of the order of the 
baryonic mass in rich clusters. For this model, the GFH peaks at
$z \sim$2 and halts at $z \sim$1.5, the metallicity peaks at $z \sim$1.5 
while evolving rather smoothly out to $z \sim$1. We consider model M2fY
as our best model.

\begin{figure}
\begin{center}
\includegraphics[width=0.35\textwidth,angle=-90]{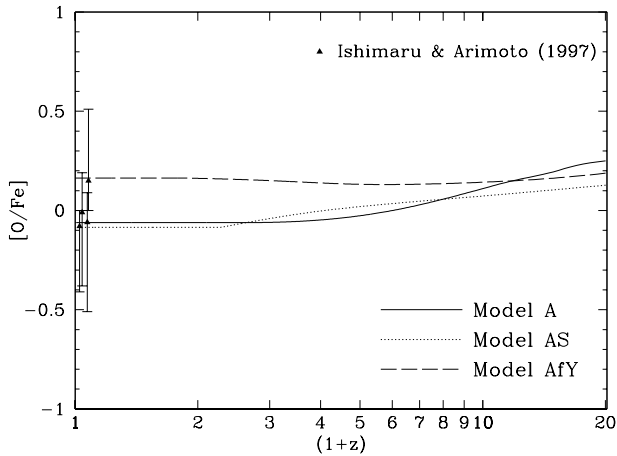}
\caption{Evolution of the [O/Fe] abundance ratio as a function of redshift
for our Models A, AS and AfY, compared to observations for low redshift 
clusters}
\label{fig:OsuFe}
\end{center}
\end{figure}

\section{Abundance ratios in the ICM}
\label{sect:abundratio}
As mentioned in the introduction, the typical [$\alpha$/Fe] abundance
ratios in the ICM are not firmly established from observations. With the
first ASCA data, it was debated if the ICM is typically $\alpha$--enriched
or if its abundances are consistent with solar ratios \citep{Mu96, IA97}.
Recent studies have revealed that abundance ratios may change in different
regions of the cluster, from roughly solar in the central regions to
$\alpha$--enriched in the outer regions \citep{F2000}. 
This might suggest competing,
or rather overlapping, extraction mechanisms of metals from galaxies:
SN~II ejecta would dominate
in GWs, while additional SN~Ia products would be
extracted from the galaxies due to ram--pressure stripping, more effective
in the central regions \citep{F2000}.

Our model at present does not deal with gradients of abundance or of
abundance ratios in the ICM, however it provides the global 
oxygen mass produced by galaxies and ejected into the ICM by GWs, 
so that we can
estimate the typical average [O/Fe] ratio in the ICM, with oxygen
being the best tracer of $\alpha$--elements. In Fig.~\ref{fig:OsuFe} 
we show the evolution
of the [O/Fe] ratio in the ICM as predicted from our reference Model~A;
[O/Fe] predictions for models M1--M2--M3 are very similar
so these models are not shown in the plot.
[O/Fe] is supersolar at very high redshifts (where the PNJ IMF
favours more massive stars and hence SN~II), decreasing 
down to marginally subsolar values.
Model~AS with the Salpeter IMF is also shown for comparison; the final [O/Fe]
value in very close to that of Model~A.

Also shown is model AfY with the ``fixed yields'' PNJ models (model M2fY is
basically indistinguishable from AfY in this plot and is not shown); 
the predicted final [O/Fe] is slightly supersolar (around +0.2 dex). Due to the
``fixed yields'' assumption in this model, [O/Fe] is roughly constant 
throughout the evolution and its value is characteristic of the PNJ models 
with $z_{for}=15$ (see \S\ref{sect:monolithic}).

Given all the above mentioned caveats about the complexity and uncertainties 
of the empirical evidence, we consider all these final values to be in broad 
agreement with observations. Notice however that all models predict a very slow
evolution (if any) in the [O/Fe] ratio, so that no sistematic trend is 
expected to be seen in for this observable until $z>3$ at the earliest.

\section{Summary and conclusions}
\label{sect:conclusions}
Galactic winds from elliptical galaxies are the most likely source of the
chemical enrichment in the ICM. In this scenario, various studies in literature
suggest that a non--standard IMF must be invoked for elliptical galaxies, 
if we are to account for the metallicity of the ICM (\S\ref{sect:gas_metal}).
\citet{C98}, \citet{C2000} calculated models of elliptical galaxies adopting 
the variable IMF by PNJ, whose behaviour in time and space is sensitive to the
physical conditions of the star--forming gas; in particular, the low--mass
cut behaves as a sort of Jeans mass (\S\ref{sect:non_stand_imf}). 
This IMF naturally predicts a lower locked--up fraction in the early galactic 
stages, especially for massive
ellipticals and/or for high redshifts of formation --- the latter feature being
related to the increasing temperature of the CMB. These galactic models
successfully reproduce a variety of observational properties of ellipticals
\citep{C98}. In \S\ref{sect:GWejecta} we show how models 
calculated with the PNJ IMF predict galaxies to eject both {\it more metals}
and {\it more gas} than ``standard'' models based on the Salpeter IMF. We 
also show that, for the more massive objects that play the main
role in the mass and metal enrichment budget of the cluster,
the characteristic IMLR of individual galaxies is much higher for models
with the PNJ IMF than with the Salpeter IMF (\S\ref{sect:GW_IMLR}).

To assess the effect of these new galactic models on the ICM, first we 
assumed all galaxies to be coeval and we integrated the corresponding ejecta
over the observed luminosity function, as in the standard ``monolithic 
approach'' (\S\ref{sect:monolithic}).
The Salpeter models fail in reproducing the observed IMLR in clusters
by an order of magnitude, while better results are obtained with the 
the PNJ IMF, especially with the models corresponding to a redshifts of 
formation $z_{for} \sim$13. We also discuss the evidence of dilution of the
galactic wind ejecta with primordial gas, introducing the concept of the
intra--cluster--mass--to--light ratio (ICMLR); for the typical luminosities and
M/L ratios of our model galaxies, a dilution of a factor 2--3 for the PNJ 
models (a factor of 6 for the Salpeter models) is necessary to recover the
typical amount of gas vs.\ galaxies observed in rich clusters.

Alternatively to the standard monolithic approach, we considered
a simple hierarchical picture in which galaxy formation extends in time, with
a characteristic galaxy mass increasing at decreasing redshift, as from
Press--Schechter theory. We developed
a toy--model following the chemical evolution of the ICM in connection
with the galaxy formation history of cluster galaxies 
(\S\ref{sect:toy-model}). 
The GFH is calibrated so as to reproduce the observed present--day galactic 
LF in clusters, and the observed amount of ICM gas (in terms of ICMLR). 
This seems to require two phases of galaxy formation: an early
phase forming dwarf galaxies populating the steep, faint tail of the LF,
followed by a more gentle GF activity peaking at $z \sim 1-2$. This scenario 
is in broad
agreement with (a) present theories and observational evidence 
about reionization and reheating of the Universe at $z \gsim 6$, and (b)
the observed trend of cosmic star formation history --- although in our cluster
models the GFH is ``anticipated'' and halts earlier
with respect to the field-based Madau--plot, since we are considering
early--type galaxies.
We mimick this ``bimodal'' galaxy formation history by means of a double 
infall prescription in our chemical model for clusters.

A satisfactory match with the observed LF can be obtained both using
galactic models with the PNJ IMF and with the Salpeter IMF. However,
models with the PNJ IMF, besides being favoured on the base of their
photometric properties \citep{C98}, provide much improved predictions 
about the metal enrichment of the ICM; Salpeter--based models 
fail in this respect (\S\ref{sect:ModelA}). Still, the PNJ models are short
of metal production by a factor of 2--3, and the conclusion does not 
depend much on the details of the GFH (\S\ref{sect:varyingGFH}). In fact,
with such extended GFHs the bulk of galaxies (in terms of mass involved)
form at redshifts $z<5$, where the PNJ models
do have a typical IMLR which is  a factor of 2--3 too low 
(\S\ref{sect:monolithic}). 

The discrepancy cannot be cured simply by increasing the (quite 
uncertain) rate of SN~Ia in the models, since the resulting [O/Fe] ratio would 
correspondingly decrease by 0.3-0.5~dex at odds with observations.
Our models presently give the correct relative contribution of SN~Ia 
and SN~II as demonstrated by the predicted [O/Fe] ratio in the ICM
( solar or slightly supersolar, \S\ref{sect:abundratio}) and
strong variations of the SN~Ia rate are not allowed within the constraint
of the observed relative abundances.

The implication is that an IMF with a more extreme behaviour 
than assumed in our galactic models is needed. Variations of the stellar 
Jeans mass with redshift induced by the CMB temperature appear
to be too mild to produce the observed metal content in clusters,
in agreement with other recent
results \citep{HF2001,Fin03}. In fact, from the monolithic approach
an IMF is required, behaving like that in the PNJ models with 
$z_{for} \sim$13--15.

We thus computed some cluster models with the IMF chosen ``ad hoc'' to induce
a metal ejection large enough to enrich the ICM to the observed
abundance level (\S\ref{sect:fixYield}). Our best model is M2fY 
(Fig.~\ref{fig:fixYield}), with a GFH peaking at $z \sim$2 and halted by 
$z \sim$1.5. This models predicts a peak in the metallicity of the ICM around 
$z \sim$1.5, decreasing at higher redshifts; at $z$=3--4 the typical 
metallicity corresponds to observed Lyman--break galaxies. In this model,
the wind--ejected gas is more than twice the mass in galaxies, and constitutes
roughly half of the ICM mass.

In general, when galactic models with the PNJ IMF are adopted, the mass 
globally ejected in GWs exceeds the mass stored in stars by a factor of 1.5--2
--- arguments in favour of large outflows of baryons from the main body of 
galaxies have been recently advanced also by \citet{Silk02}.
Although this is not enough to account for the total mass in the ICM, it is
nevertheless a non--negligible fraction of the intra--cluster gas, especially
for less rich clusters. This suggestion is reinforced by recent results
obtained with dynamical models of galaxy formation \citep{Carr01,Ch02},
showing that the process of galaxy formation is so unefficient that
roughly 75\% of the gas initially available is ``wasted'' and expelled,
while only a 25\% remains locked in the final galaxy. These dynamical models,
{\it even adopting a constant IMF}, suggest a typical ratio of 1:3 between 
the matter stored in galaxies and that
re-ejected in the inter--galactic, or intra--cluster, medium. It is suggestive
to speculate that 3:1 or so could set a minimal value for the gas--to--galaxy
mass ratio in clusters; this value could then be typical of 
poor clusters, whose ICM would be dominated by gas of galactic
origin.
This is also in line with the observational evidence that 
the energetics of the ICM in the less rich and luminous clusters are dominated
by non--gravitational effects, most likely related to the energy feed-back
from cluster galaxies.

\acknowledgements
We would like to thank our referee, Prof. A.~Renzini, for very helpful critical remarks.\\
We benefitted also from discussions with S.~Andreon, A.~Finoguenov, M.~G\"otz,
E.~Pignatelli and J.~Sommer--Larsen. 
LP acknowledges kind hospitality from the Astronomy Department 
in Padova and from the Observatory of Helsinki upon various visits.\\
This study was financed by the Italian MIUR and University of Padova under the special contract ``Formation and evolution of Elliptical Galaxies'' and by 
the Danmarks Grundforskningsfond (through its support to TAC).

\bibliographystyle{aa}
\bibliography{biblio}

\end{document}